\documentclass[twocolumn]{aastex631}

\shorttitle{Roman Exomoons}
\shortauthors{Lastovka et al.}

\graphicspath{{./}{figures/}}

\begin{document}

\title{Predictions of the Nancy Grace Roman Space Telescope Galactic Exoplanet Survey. III. Detectability of Giant Exomoons of Wide Separation Giant Planets}

\author[0009-0009-4693-7783]{Matthew Lastovka}
\affiliation{Department of Astronomy, University of Maryland, College Park, MD 20742, USA}
\affiliation{Department of Astronomy, The Ohio State University,
Columbus, OH 43210, USA}

\author[0000-0003-0395-9869]{B. Scott Gaudi}
\affiliation{Department of Astronomy, The Ohio State University,
Columbus, OH 43210, USA}

\author[0000-0001-9397-4768]{Samson A. Johnson}
\affiliation{Department of Astronomy, The Ohio State University,
Columbus, OH 43210, USA}

\author[0000-0001-7506-5640]{Matthew T. Penny}
\affiliation{Department of Physics and Astronomy, Louisiana State University, Baton Rouge, LA 70803, USA}

\author[0000-0002-1743-4468]{Eamonn Kerins}
\affiliation{Jodrell Bank Centre for Astrophysics, Alan Turing Building, University of Manchester, Manchester M13 9PL, UK}

\author[0000-0001-5069-319X]{Nicholas J. Rattenbury}
\affiliation{Department of Physics, University of Auckland, Private Bag 92019, Auckland, New Zealand}

%%%%%%%%%%%%%%%%%%%%%%%%%%%%%%%%%%%%%%%%%%%%%%%%%%%%%%%%%%%%%%%%%%%%%%%%%%%%%%%%
\begin{abstract}

The \textit{Nancy Grace Roman Space Telescope} (Roman) will conduct a Galactic Exoplanet Survey (RGES) to discover bound and free-floating exoplanets using gravitational microlensing. Roman should be sensitive to lenses with mass down to $\sim 0.02 \; M_{\oplus}$, or roughly the mass of Ganymede. Thus the detection of moons with masses similar to the giant moons in our Solar System is possible with Roman. Measuring the demographics of exomoons will provide constraints on both moon and planet formation.  We conduct simulations of Roman microlensing events to determine the effects of exomoons on microlensing light curves, and whether these effects are detectable with Roman. We focus on giant planets from 30 $M_{\oplus}$ to 10 $M_{Jup}$ on orbits from 0.3 to 30 AU, and assume that each planet is orbited by a moon with moon-planet mass ratio from $10^{-4}$ to $10^{-2}$ and separations from 0.1 to 0.5 planet Hill radii. We find that Roman is sensitive to exomoons, although the number of expected detections is only of order one over the duration of the survey, unless exomoons are more common or massive than we assumed.  We argue that changes in the survey strategy, in particular focusing on a few fields with higher cadence, may allow for the detection of more exomoons with Roman.  Regardless, the ability to detect exomoons reinforces the need to develop robust methods for modeling triple lens microlensing events to fully utilize the capabilities of Roman. 

\end{abstract}


%%%%%%%%%%%%%%%%%%%%%%%%%%%%%%%%%%%%%%%%%%%%%%%%%%%%%%%%%%%%%%%%%%%%%%%%%%%%%%%%
\section{Introduction} \label{sec:intro}

Despite the discovery of nearly 6000 exoplanets to date\footnote{See https://exoplanetarchive.ipac.caltech.edu/}, little is understood about the populations of satellites these systems may harbor.
%To date, over 5600 exoplanets have been discovered\footnote{See https://exoplanetarchive.ipac.caltech.edu/}.  These systems span a wide range of host star and planet properties (see, e.g., \citealt{Zhu_2021}), and include populations of planets, such as ``Hot Jupiters'' and ``Super-Earths", that have no analogs in our Solar System. 
%It has been a struggle to explain this diversity of exoplanetary systems, and consequently, planet formation remains poorly understood.
In our Solar System, moons appear to be a common by-product of planet formation. Nearly 300 moons are known to orbit the planets in our Solar System, and thus moons outnumber planets by nearly forty to one.  Of these, seven have masses $\ge 10^{-3}~M_\oplus$.  Notably, all four giant planets host relatively massive moons with orbits that are roughly circular and coplanar to the equator of the planet, likely indicating that they were formed concurrently with the planet in a circumplanetary disk \citep{Canup_2002,Canup_2006,Sasaki_2010,Ogihara_2012,Batygin_2020,Cilibrasi_2018,Shibaike_2019,Ronnet_2020,Heller_2014}.

Given the ubiquity of moons orbiting the giant planets in our Solar System, it is interesting to consider whether exomoons are also common. Determining the demographics of exomoons can provide insights into the physical processes that govern moon and planet formation.  Furthermore, the presence of an exomoon may impact the habitability of its host planet (e.g., \citealt{Laskar_1993}), and exomoons may themselves be habitable \citep{Williams_1997,Heller_2013,Heller_2014,Kaltenegger_2010}.  

A number of ways have been proposed to detect exomoons, however there have been no confirmed detections to date. Possible detection methods include direct imaging of exomoons, observing the transit of a moon on a directly imaged exoplanet, the Rossiter-McLaughlin effect, microlensing, transit timing variations, and more\footnote{See \citet{Heller_2014} for a review of the formation, habitability, and detection of exomoons.}. Two exomoon candidates, Kepler-1625bi and Kepler-1708bi, have been identified by observing a small dip in a light curve immediately before or after the transit of their respective planet \citep{Kipping_2022,Teachey_2018}. 

In this work, we will focus on gravitational microlensing as a method to detect exomoons. This technique is effective because it is directly sensitive to the mass of the lens object \citep{Mao_1991,Gould_1992} and can produce relatively strong signals, even for bodies much less massive than Earth \citep{Bennett:1996}, as long as sufficiently small source stars are monitored.

\citet{Bennett:2002} were the first to propose a space-based microlensing survey and demonstrated its potential. They noted that such a survey could detect moons similar to the giant moons in our solar system. Since then, several researchers have explored the characteristics and detectability of light curve perturbations caused by moons orbiting both bound and free-floating exoplanets \cite{Han:2002,Han:2008,Chung:2016,Sajadian:2023,Liebig_2010,Koshimoto2023,Mroz2019}.

We motivate our focus on moons orbiting planets with wide separations from their stars detected by Roman here. We note that the signals due to moons that are widely separated from their host planets generally do not decrease in magnitude with increased distance from the planet \citep{Han:2008}.  This is because the moon signals become dominated by the gravitational influence of the star on the moon\footnote{Formally, the moon signals become dominated by the gravitational 'shear' of the star at the location of the moon.  See, e.g., \cite{Mao:92} and Section \ref{sec:general} for a discussion of gravitational shear.} rather than the gravitational influence of the planet on the moon. Furthermore, the sensitivity of microlensing to planets outside the snow line allows one to probe regions of planetary systems more likely to host moons, or at least moons formed in circumplanetary disks \citep{Heller_2014}. Finally, all other parameters being equal, orbits of moons around wide-separation planets also tend to be more stable \citep{Barnes:2002}, and the range of semimajor axes with the greatest stability is fortuitously located close to the region where moons are most readily detectable by microlensing.  

The prospect of detecting exomoons using microlensing has been greatly enhanced by the impending launch of the \textit{Nancy Grace Roman Space Telescope} (Roman) \citep{Spergel:2015}, currently due for launch in late 2026. Roman was chosen as the top-priority large space mission of the 2010 Astronomy and Astrophysics Decadal Survey.  See \citet{Akeson:2019} for an overview. Roman will have three Core Community Surveys: a High Latitude Wide Area Survey that will use imaging and spectroscopy to measure cosmological parameters using galaxy clustering and weak lensing \citep{Spergel:2015}, a High-Latitude Time Domain Survey that will image tens of square degrees on a few-day cadence to discover and characterize thousands of Type 1a Supernovae \citep{Hounsell:2018}, and a Galactic Bulge Time Domain Survey (RGBTDS) that will monitor a few square degrees on a nominal 15 minute cadence to discovery thousands of exoplanets via microlensing and measure their demographics. It will also have a Galactic Plane Survey that will survey $\sim 1000$ square degrees of the bulge and inner Galactic disk to study the stellar populations of our inner Galaxy \citep{Paladini:2023}. In addition, at least 25\% of the 5-year prime mission will be devoted to a competed General Observer Program.  Finally, Roman will be equipped with a coronagraph designed for $>10^8$ starlight suppression to directly image planets in reflected visible light \citep{Bailey:2023}.

The survey strategy and parameters of the RGBTDS we consider here is that which was studied by \citet{Penny_2019}.  It is a $\sim 432$ day survey comprised of 6 seasons of 72 days each. A total of $\sim 2$ square degrees will be monitored every $\sim 15$ minutes in a wide, $0.927-2~{\mu}m$ F146 filter with lower cadence observations in one or two other filters, to be determined.  \citet{Penny_2019} and \citet{Johnson_2020} demonstrated that this survey should detect thousands of bound and free-floating exoplanets.  In particular, it should be sensitive to planets with masses greater than $\sim 0.02 M_\oplus$, or about the mass of Ganymede. This raises the prospect that microlensing effects due to exomoons with masses of Ganymede and greater could be detectable by Roman. Furthermore, Roman has significant sensitivity to massive planets on relatively wide orbits, which are likely where moons are most readily detectable, as we explain below.  

Recently, after the initial submission of this manuscript and during the refereeing process, the Roman Observations Time Allocation Committee recommended a prioritized design for the RGBTDS that differs slightly from the one considered here \footnote{\url{https://roman.gsfc.nasa.gov/science/ccs/ROTAC-Report-20250424-v1.pdf}}. Of particular relevance to the results of this paper, they recommended five primary microlensing survey fields (plus one Galactic center field), each monitored during six high-cadence seasons of 72 days. During these seasons, each field is observed with a cadence of 12.1 minutes. The total survey duration is 440 days.

This updated survey strategy is expected to result in a slightly higher exomoon yield than the one presented here. However, given the small number of expected exomoon detections, Poisson fluctuations in the actual yield are likely to be larger than this difference. As a result—and due to the significant computational cost of repeating the simulations—we chose not to attempt a revised yield estimate based on the new survey strategy.

Here we quantify Roman's sensitivity to moons with low moon/planet mass ratios ($\sim 10^{-2}-10^{-4}$), focusing on companions to wide-separation planets detected by Roman Galactic Exoplanet Survey (RGES) component of the Roman Galactic Bulge Time Domain Survey.  The structure of the paper is as follows. Section \ref{sec:micolens} reviews microlensing by one, two, and three bodies, and Section \ref{subsec:rayshoot} describes our inverse ray-shooting algorithm for computing light curves of triple lenses.  Section \ref{sec:general} describes general considerations of microlensing by planets with moons in order to anticipate some of our results. Section \ref{sec:distribution} motivates and describes our assumed distributions for the planet and moon properties. Section \ref{sec:romansim} describes our simulations of the RGES survey that we use to assess the yield of exomoons, and Section \ref{sec:results} presents our results. In Section \ref{sec:discussion}, we discuss the two primary channels by which exomoons can be detected, the impact of the sampling rate on the yield, the effect of multiple moons on the planet yield, and the error induced by the fact that we do not consider the detectability of moons in resonant planetary events. Finally, Section \ref{sec:conclusions} presents our conclusions.

\section{Microlensing by One, Two, and Three Masses} 
\label{sec:micolens}

A microlensing event occurs when a foreground object (the lens) passes very close to our line-of-sight to a background star (the source), where `very close' is typically defined to be within one angular Einstein ring $\theta_{\rm E}$ radius of the lens, 
\begin{equation}
    \theta_{\rm E} \equiv \sqrt{\kappa M \pi_{\rm rel}}\label{eqn:thetae}. 
\end{equation}
Here $M$ is the mass of the lens, $\pi_{\rm rel} \equiv {\rm AU} \left( 1/D_l - 1/D_s \right) = \pi_l-\pi_s$ is the relative lens-source parallax, with $\pi_l$, $D_l$ and $\pi_s$, $D_s$ being the parallax and distance to the lens and source, respectively. The constant $\kappa \equiv 4G/c^2{\rm AU} \simeq 8.14~{\rm mas}\,M_\odot^{-1}$.  Note that $\theta_{\rm E}$ is the radius of the ring that is created with the source and lens are perfectly aligned.  It is also useful to define the physical size of the Einstein ring radius at the distance to the lens, $R_{\rm E}=D_l \theta_{\rm E}$. Generally $\theta_{\rm E}$ and $R_{\rm E}$ are used as the the fiducial angular scale and unit of length in microlensing.   When this transient alignment occurs, the light from the background source gets split into multiple images.  The images can also be magnified or demagnified relative to the unlensed source.

For $N_l$ point masses, the positions of these images are given by the lens equation
\begin{equation} \label{eq:lens}
    \zeta = z - \sum_{i}^{N_l} \frac{\epsilon_i}{\overline{z} - \overline{z}_{m,i}}
\end{equation}
where $\zeta$ and $z$ are the source and image positions, respectively, expressed in complex form, 
$z_{m,i}$ is the location of mass $i$, and $\epsilon_i \equiv m_i/M$, where $M$ is the combined mass of the lenses. Both $z$ and $\zeta$ are in units of the Einstein ring radius for the total mass of the lens. 

Microlensing conserves surface brightness. However, the total area of the images is not necessarily equal to the area of the source, causing the magnification of the source. The most straightforward way of calculating the magnification is to invert the lens equation (Equation \ref{eq:lens}) and to determine the positions of the images as a function of the locations of the source and lenses. This can be done analytically for the single-lens case.  For the binary lens case, Equation \ref{eq:lens} is equivalent to a fifth-order polynomial in $z$, and which must be solved numerically.  Note that not all solutions to the polynomial are solutions to Equation \ref{eq:lens}. Assuming a point source, the magnification of each image is given by the determinant of the Jacobian of the mapping in Equation \ref{eq:lens}, evaluated at that image position.  The total magnification of the source is then just the sum of the (absolute values) of the magnifications of each image. See \citet{Gaudi:2012} for more discussion and the relevant formulae.  

For a single lens, two images are created, one on the same side as the source from the lens and outside the Einstein ring and one on the opposite side of the source from the lens and inside the Einstein ring. The magnifications of these images are,
\begin{equation} \label{eq:mag0twoimages}
    A_{\pm} = \frac{1}{2}\frac{u^2+2}{u\sqrt{u^2+4}} \pm 1,
\end{equation}
where $u$ is the angular separation between the source and lens in units of $\theta_{\rm E}$.  These images are separated by $\sim 2 \theta_{\rm E}$ when the magnification is significant, and are not resolved with current technology for most events.  The characteristic timescale for microlensing events is the Einstein ring crossing time,
\begin{equation}
t_{\rm E}\equiv \frac{\theta_{\rm E}}{\mu_{\rm rel}},
\label{eqn:tE}
\end{equation}
where $\mu_{\rm rel} = \mu_{l}-\mu_{s}$ is the relative proper motion between the lens and the source.  

\begin{figure}
    \centering
    \includegraphics[width=0.9\linewidth]{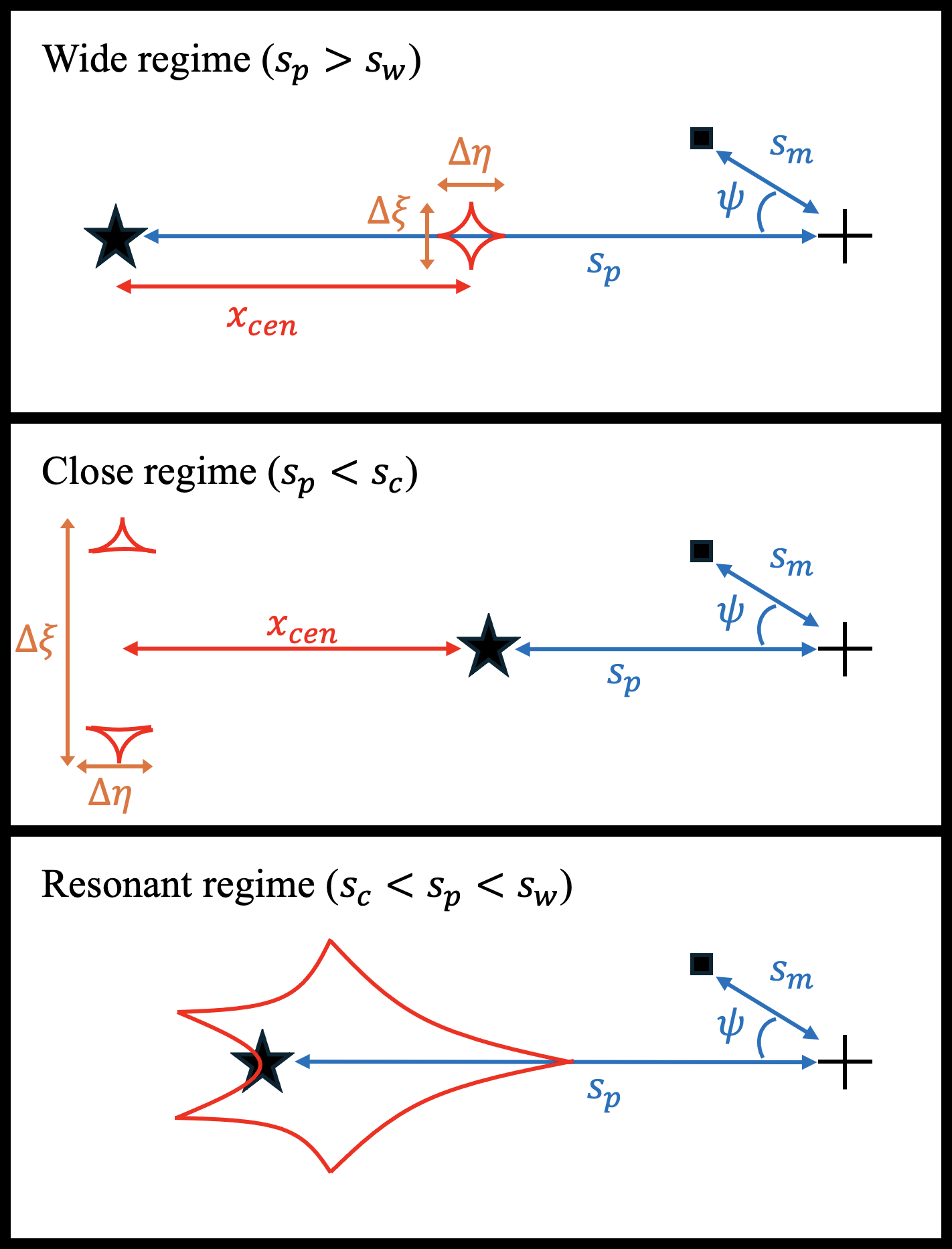}
    \caption{Geometry of lens systems and their caustics. Each panel shows a different regime of caustic structure, defined using the criteria from Equation \ref{eq:caustic_topology}. The star, plus, and square indicate the positions of the star, planet, and moon, respectively. $s_p$ and $s_m$ indicate the distances between the star and planet, and the moon and planet, respectively. $\psi$ indicates the orientation of the moon relative to the planet-star axis. The red lines show the location and shape of the planetary caustic(s), or, in the case of the resonant regime, the resonant caustics.  For the close and wide regimes, $x_{cen}$ shows the distance between the star and the center of the caustic. $\Delta\xi$ and $\Delta\eta$ are the vertical and horizontal widths of the planetary caustic, respectively. The cross section of the planetary caustic, $s_{caus}$, is the geometric mean of $\Delta\eta$ and $\Delta\xi$. Note: this figure is not drawn to scale.}
    \label{fig:geometry}
\end{figure}

An important property of gravitational microlenses is the existence of caustics.  Caustics are defined as the loci of points in the source plane where the magnification of a point source is formally infinite, and where the number of real images change by two or more when the source crosses a caustic.   
For a binary lens, there are either one, two, or three closed caustic curves \citep{Schneider:1986}, depending on $s_p$ and $q_p$, where $s_p$ is the instantaneous separation between the planet and star in units of $\theta_{\rm E}$ normalized to the total mass of the lens and $q_p$ is the planet-star mass ratio.  For $q_p\ll 1$, the boundaries between these three caustic topologies are given by  (\citealt{Schneider:1986,Dominik:1999}):  
\begin{equation} \label{eq:caustic_topology}
 s_c \simeq 1-\frac{3}{4}q_p^{1/3{}},\;\;\;\;\;\; s_w \simeq 1+\frac{3}{2}q_p^{1/3}
\end{equation}
In the regime of $q_p\ll 1$ and for the close topology ($s_p<s_c$), there are three caustics, a central caustic located near the star, and two triangular-shaped planetary caustics straddling the planet-star axis.  For the resonant topology ($s_c<s_p<s_w$), there is one relatively large caustic near the star.  Finally, for the wide topology ($s_p>s_w$), there are two caustics, a central caustic located near the star, and a planetary caustic.  For the close and wide topologies, the planetary caustics are located at a distance of 
\begin{equation}\label{eq:caustic_eqn}
    x_{cen}=s_p-s_p^{-1}
\end{equation}
from the star along the planet-star axis.  For reasons that we explain in the next section, we will be primarily focusing on cases where the moon perturbs the wide planetary caustics, although we will also be considering cases where the moon perturbs the close planetary caustics.  We will not be exploring the detectability of moons orbiting planets in the resonant topology, due to computational difficulties. See Figure \ref{fig:geometry} for a visualization of the three caustic topologies.

Systems where a moon orbits a bound planet are triple lenses ($N_l=3$), for which Equation \ref{eq:lens} is equivalent to a tenth-order polynomial in $z$. As with a binary lens, this polynomial must be solved numerically to determine the image positions for a given source position and lens positions, and the magnification of each image for a point source is the determinant of the Jacobian of the mapping in Equation \ref{eq:lens} evaluated at that image position.  

The approximation of a point source breaks down when the source approaches a caustic.   In this case, the finite size of the source must be considered when computing the magnification.  There are several methods for computing microlensing light curves for finite sources, including integrating over boundaries of the images of the source and then computing the area of each image via Green's theorem \citep{Gould:1997,Bozza:2010}, or by image-centered \citep{Bennett:1996,Bennett:2010} or global \citep{Wambsganss:1997} inverse ray shooting.

There are robust, publicly available routines for computing light curves considering finite source effects using these techniques for binary lenses (e.g., \citealt{Bozza:2018,Bachelet:2017,Poleski:2019,Bennet_2010}). However, up until after we completed the primary computations for this paper, there existed only one analogous, publicly available code for triple lenses \citep{Kuang_2021}\footnote{Recently, the code {\tt VBMicrolensing} was made publicly available (\citealt{Bozza:2025}).  This code computes light curves with finite source effects for an arbitrary number of lenses.}. We found that this code produced numerical errors for the very low mass ratios ($\sim 10^{-8}$) we consider here, presumably due to numerical noise in the image positions computed by inverting the tenth-order polynomial.  We therefore chose to develop our own image-centered inverse ray shooting code to compute the magnification of a triple lens consisting of the following hierarchy: $q_m q_p\ll q_p \ll 1$, where $q_m \equiv M_m/M_p$, $q_p \equiv M_p/M_*$, and $M_m$, $M_p$, and $M_*$ are the masses of the moon, planet, and star, respectively.

The magnification of any source by a moon-planet-star system depends on the location of the source with respect to the origin of the coordinate system and the parameters $q_m$, $q_p$, $s_m$, $s_p$, and $\Psi$.  Here $s_m$ is the instantaneous moon-planet separation in units of the Einstein ring of the planet $\theta_{{\rm E},p} \equiv q_p^{1/2} \theta_{\rm E}$, and $\Psi$ is the angle of the moon-planet separation defined clockwise relative to the planet-star separation, such that for $\Psi=0$ the moon is directly between the planet and the star. See Figure \ref{fig:geometry} for a visualization of the system geometry.  This assumes a point source; in the case of a finite source, the magnification also depends on $\rho_*\equiv \theta_*/\theta_{\rm E}$, the angular radius of the source $\theta_*$ in units of $\theta_{\rm E}$.  

\subsection{Inverse Ray-Shooting} \label{subsec:rayshoot}

The inverse ray-shooting technique takes advantage of the fact that the surface brightnesses of the images are smooth and continuous, and magnification is simply the total area of the images divided by the area of the source.  Thus, instead of tracing light rays from the background star to the observer, rays are traced from the observer back to the source plane. The effect of each lens on the ray is calculated, and then each ray is collected into pixels in the source plane. The result is a map with the number of rays in each pixel proportional to the magnification of a background source with the size of the pixel at that location. This technique allows us to easily account for finite source effects, and adding additional lenses does not significantly increase computational complexity.

One drawback of inverse ray-shooting is that it is a computationally expensive process. This is particularly true when one of the images becomes large, e.g., when the magnification of the image is much greater than unity (typically $\gtrsim 100$).  Furthermore, for high-magnification events, the moon will perturb both images, further increasing the computational expense as both images must be covered with rays.  We therefore chose to focus on cases where the moon is perturbing the planetary caustic in the close or wide topology planetary configurations.  We therefore do not consider cases where the moon is perturbing a resonant caustic created by the planet, as these would require shooting many rays over a large area of the image plane. The boundaries in $s_p$ between the close/resonant and resonant/wide topologies are given in Equation \ref{eq:caustic_topology}; we do not consider the detectability of moons orbiting planets with $s_c \ge s_p \ge s_w$.  We note that, because the mapping between the planet's semimajor axis $a_p$ and $s_p$ depends on the $\theta_{\rm E}$ as well as the planet's phase and orbital elements, the excluded range in $s_p$ does not correspond to a fixed range in $a_p$.  Furthermore, because the range of $s_p$ spanned by the resonant topology shrinks as $q_p^{1/3}$ \citep{Gaudi_2010}, this becomes less important for lower planet/star mass ratios.   

Planetary caustics arise when the planet perturbs one of the two images created by its host star.  In this case, the other image is essentially unperturbed.  Similarly, moons orbiting planets creating planetary caustics will only significantly perturb that image created by the star and will have a negligible effect on the other image.
%\footnote{Equivalently, one can imagine that the moon creates a significant additional perturbation to one of the two or four images the planet creates when it perturbs one of the two images of created by the host star.}.
Therefore, to further improve efficiency, we only calculate magnification maps for the region of the source plane that encloses the caustics produced by the planet and moon. First, we center the ray shooting region at the location of $x_{cen}$ of the planetary caustics, where $x_{cen}$ is given by Equation \ref{eq:caustic_eqn}. We define the size of this region as four times the Einstein ring radius of the planet $\theta_{{\rm E},p}$. Then, we translate this box to the lens plane to obtain the region of the lens plane over which we will apply ray-shooting. The size of the ray-shooting region in the lens plane is scaled by the magnification in that region, which we take to be the single-lens magnification at the center of the caustic, which can be calculated analytically using Equation~\ref{eq:mag0twoimages}. 

\begin{figure}
    \centering
    \includegraphics[width=\linewidth]{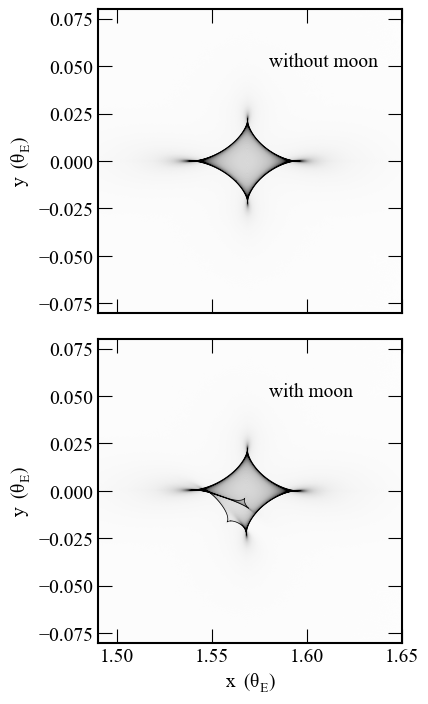}
    \caption{An example magnification map generated using inverse ray-shooting and the following parameters: $q_p=0.0026$, $q_m=0.01$, $s_p=2.058$, $s_m=0.9648$,
    $\Psi=43^\circ$. On this scale, darker color indicates higher magnification. The moon introduces additional structure to the caustic that is not present with just a planet.}
    \label{fig:magnification_map}
\end{figure}

Having defined the ray-shooting region of the lens plane, we shoot rays uniformly in a grid in that region.  For this work, we shoot $5.8564 \times 10^8$ rays. We then calculate how each ray is deflected by the lenses using the lens equation \ref{eq:lens}. All the rays are then collected and 
binned into pixels in the source plane. The magnification in each pixel is then proportional to the number of rays landing in each pixel. To convert to magnification, we use the relationship
\begin{equation}
    A_{i,j}=\frac{N_{i,j}}{A_s  \sigma_l }
\end{equation}
where $N_{i,j}$ is the number of rays in each pixel, $A_s$ is the area of the pixel, and $\sigma_l$ is the number density of rays shot in the lens plane. Since we only shoot over a small region of the lens plane, this does not represent the total magnification,  only the magnification of the perturbed image. As discussed above, this magnification does not account for the second image created by the star that is not perturbed by the planet or moon. To account for this non-perturbed image, we add the single lens magnification of that image, which can be calculated analytically using Equation \ref{eq:mag0twoimages}, where $+$ for the `major' image outside the Einstein ring and $-$ for the `minor' image inside the Einstein ring. For the close topology ($s < s_c$), the image perturbed by the moon is the minor image, so we add the single lens magnification for the major image to the magnification. Conversely, for the wide topology, we add the minor image magnification to the magnification. 

This process results in a pixelated magnification map: the magnification of a two-dimensional grid of pixels for a given $q_p$, $s_p$, $q_m$, $s_m$, and $\Psi$. Figure \ref{fig:magnification_map} shows an example magnification map using the following parameters: $q_p = 0.0026$, $q_m=0.01$, $s_p=2.058$, $s_m=0.9648$, $\Psi=43^{\circ}$.

A microlensing light curve is simply a one-dimensional interpolation through this pixelated magnification map, convolved with the source surface brightness profile.  One advantage to inverse ray-shooting is that it can easily include finite source effects. We average over the magnification values of all pixels within the source to get the total magnification of the source. This is trivial for pixels that are entirely inside of the source boundary. However, some pixels only have a fraction of their area within the source, with the rest outside. Here, we count any pixel that has more than half its area inside of the source and we reject any pixel with more than half of its area outside of the source. This is a reasonable approximation as long as the size of each pixel is much smaller than the area of the source. We set the pixel size for each case to ensure a constant ratio between the pixel size and source size. We set the pixel width for all pixels to be $0.055 \rho_*$, making the area of each pixel 1040 times smaller than the area of the source.
%We choose a pixel size that is at least about 1000 times smaller than the area of the source.

We assume a uniform surface brightness for the source.  True sources are limb-darkened, which will change the detailed shape of the light curve near areas where the gradient of the magnification of the source across the source is large.  In general, the features arising from a limb-darkened source will be `rounder' than those due to a uniform source.  However, for the relatively red Roman filter we consider here (W149), the limb darkening should be relatively modest, and so our approximation of a uniform source will not change our conclusions qualitatively.   

Since we only create a magnification map for a small region of the source plane around the planetary caustics, only a fraction of the light curve is computed. To compute the rest of the light curve for which the source is outside of the ray-shooting region, we use {\tt MulensModel} \citep{Poleski:2019}. {\tt MulensModel} uses the {\tt VBBinaryLensing} algorithm \citep{Bozza:2018} to calculate light curves\footnote{Recently, {\tt VBBinaryLensing} was updated to {\tt VBMicroLensing} \cite{Bozza:2025}, which has since been encorporated into {\tt MulensModel}.  We used the implementation of {\tt MulensModel} with {\tt VBBinaryLensing} in this paper.  We expect the difference between these two implementations to be negligible.}, which itself uses the advanced contour integration method. This method generally has negligible numerical noise compared to the photometric uncertainties of the events simulated here. We assume that the moon does not have any effect on the light curve outside of the ray-shooting region, which allows us to utilize the efficient computation of binary lens light curves that {\tt MulensModel} offers.

\subsection{Verification of the Ray-Shooting Algorithm}\label{subsec:verify}

\begin{figure}
    \centering
    \includegraphics[width=0.48\textwidth]{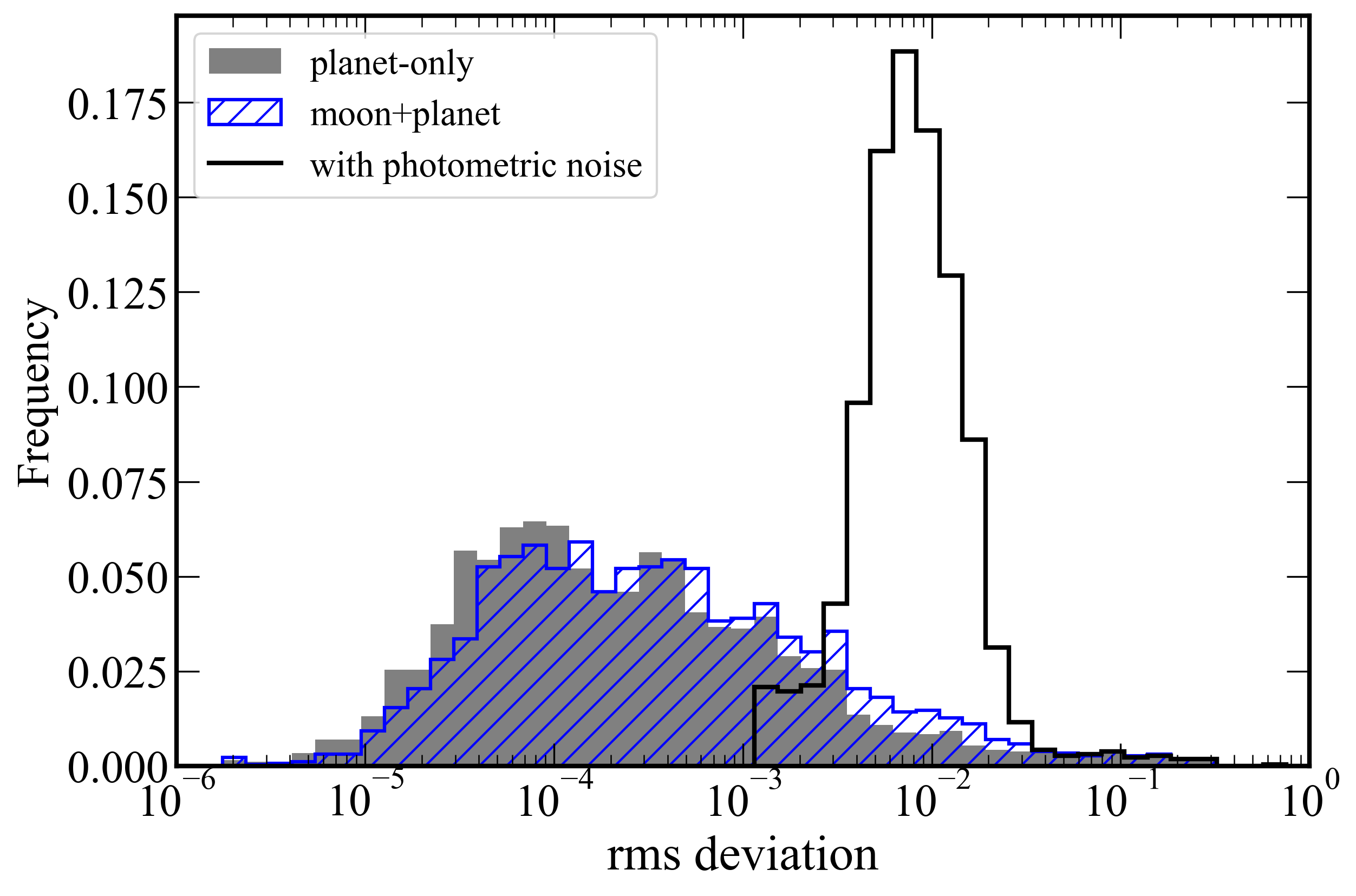}
    \caption{A histogram of the root-mean-square (rms) deviation between light curves computed using our inverse ray-shooting algorithm and light curves computed using {\tt MulensModel} \citep{Poleski:2019}. We use a subset of 2590 light curves computed as part of the simulations described in this work. For the rms calculation, we only use the region of the light curve calculated using ray-shooting and exclude the parts calculated with {\tt MulensModel}.
    The gray-filled histogram shows the rms between ray-shooting light curves with only a planet and a {\tt MulensModel} light curve calculated with the same parameters. The blue hatched histogram uses light curves that include a moon compared to the same {\tt MulensModel} light curve. The open histogram shows the rms between planet-only ray shooting light curves (no moon), including expected photometric noise (see Section \ref{sec:romansim}), compared to the same {\tt MulensModel} light curve. The blue histogram with the moon is essentially the gray (planet-only) histogram shifted to larger deviations, resulting in a slight excess compared to the planet-only histogram at high rms that comes from the perturbations due to the moons.}
    \label{fig:verify}
\end{figure}

To verify that our inverse ray-shooting algorithm is accurately computing microlensing light curves, we take a sample of microlensing events with a detectable planetary signal and compute the light curves using our algorithm, then compare with light curves calculated using {\tt MulensModel}. We use a sample of 2590 microlensing events computed as part of the simulations described in this work. For each event, we calculate three light curves using our inverse ray-shooting method: (1) the light curve with only a planet, (2) the light curve including a moon (which should be the same as (1) except for the perturbation from the moon, and (3) the planet-only light curve including the expected Roman photometric noise (see Section \ref{sec:romansim}). (1) and (2) show the numerical noise from our ray-shooting calculations, which we compare with the photometric noise from (3).

We calculate the root-mean-square (rms) deviation between each of these light curves and a light curve calculated using {\tt MulensModel} and show the results as histograms in Figure \ref{fig:verify}. To calculate the rms, we only include parts of the light curves calculated with ray-shooting, and only consider those parts of the light curves with more than 5 data points. 

We see that the histogram of rms values for light curves with the moon (blue) is nearly identical to the histogram with the planet only (gray). This is not surprising, as the noise in the inverse ray shooting algorithm arises from Poisson fluctuations due to the finite number of rays that ‘land’ in the source plane. The magnitude of these fluctuations depends on the density of rays shot in the image plane and the local magnification.  As we shoot the same number of rays in both the triple and binary lens cases, and the magnifications in the two cases are similar, we expect the noise to be similar. This is in contrast to some algorithms that rely on solving the complex polynomial that is derived from the lens equation, which are inherently noisier for higher-order polynomials due to how the roots are found.  We note that the triple-to-binary lens histogram is essentially the binary-to-binary histogram shifted to larger deviations, resulting in a slight excess compared to the binary-to-binary histogram at high rms that comes from the perturbations due to the moons themselves.

For the planet-only histogram, we find a median rms value of $1.9\times10^{-4}$.
For the histogram including photometric noise, we find a median rms of $7.6\times10^{-3}$, which is a factor of $\sim 40$ times larger than the median numerical rms. We therefore conclude that the numerical noise in our computed light curve is negligible for the majority of cases. 

We further eliminated some potential false positives during our search for detected moons by computing two light curves for each event, one that includes the moon and one that does not. We then fit both light curves to a planet-only model computed using {\tt MulensModel}, and use the difference in $\chi^2$ of these two fits to determine the significance of the moon perturbation (see Section \ref{sec:romansim} for more details). By calculating both planet and planet+moon perturbations using inverse ray-shooting, but fitting both to an essentially noise-free model, we account for the numerical noise due to the inverse ray-shooting method.  Since Figure \ref{fig:verify} demonstrates that the noise impacts the triple lens and binary lens light curve computations by approximately the same amount, this procedure should eliminate any potential false positives.

Finally, to ensure that there were no false positives among the simulated detections, we visually inspected all initial detections to ensure that none were due to numerical errors.  During this process, we found and removed a small number (roughly 15) of formally significant detections that were due to the breakdown of our assumption that we were only sampling one of the two images created by primary. 

%We note that $\sim 20$\% of our planet-only light curves simulated using our inverse ray shooting algorithm have an rms relative to {\tt MulensModel} calculation that is greater than the minimum rms due to photometric noise of $1\times10^{-3}$. It is important to note that for most of these cases, the actual photometric noise will be larger than $10^{-3}$ and the numerical noise will still be comparatively small. By computing two light curves as described above, any cases where the numerical noise does become important will not be flagged as detections. However, we found that there are a small subset of cases with rms $\sim 10^{-1}$. Most of these cases are due to a numerical issue caused by the breakdown of our assumption that our ray-shooting code only calculates the magnification from one of the images. This is a good assumption for the wide and close caustic regimes, where the planet only perturbs one of the images. This assumption breaks down, however, for cases in and near the resonant regime, creating nonphysical discontinuities in the light curve. To account for this, we visually inspect all light curves flagged as having detectable moons by our simulations and remove a small number (roughly 15) suffering from this issue.

\section{General Considerations}\label{sec:general}

Here we are considering the hierarchical system of three bodies with $q_m q_p \ll q_p \ll 1$.  A moon is a body that orbits within the sphere of gravitational influence of the planet, which is typically called the Hill sphere and has a radius
\begin{equation}
a_{\rm Hill} = a_p \left(\frac{q_p}{3}\right)^{1/3},
\label{eqn:hillradius}
\end{equation}
which can be derived for the circular restricted three-body problem by equating the gravitational and centrifugal forces on the third, test body and assuming $q_p\ll 1$.  

Bodies with semimajor axes with $a \le a_{\rm Hill}$ are not necessarily stable.  For example, using a suite of numerical simulations, \citet{Domingos:2006} find that satellites on prograde orbits with $a \le 0.4895 a_{\rm Hill}$ are stable (see also \citealt{Holman:1999}), while those on retrograde orbits with $a\le 0.9303 a_{\rm Hill}$ are stable. Satellites with orbits larger than this critical $a$ are in unstable orbits. Furthermore, tidal interactions between the star, planet, and moon can also serve to destabilize moons on long timescales \cite{Barnes:2002}.  All else fixed, moons can be stable for longer periods of time when orbiting planets with larger semimajor axes. Here we consider moons with semimajor axes in the range $0.1 \le a/a_{\rm Hill}\le 0.5$.

Although there have been a number of studies that have examined the detectability of moons with microlensing \citep{Han:2002,Han:2008,Chung:2016,Sajadian:2023,Liebig_2010}, the phenomenology of microlensing light curves due to moon-planet-star systems remains poorly understood. To date, there has been no comprehensive study of how the morphology and magnitude of perturbations due to moons depend on $q_p$, $s_p$, $q_m$, $s_m$, $\Psi$, and $\rho_*$.  Indeed, although the topologies of the caustics for 3-body microlenses have been systematically explored \citep{Danek:2015,Danek:2019}, with attention paid to the hierarchical case of $q_m q_p \ll q_p \ll 1$ relevant to our study, these studies have been highly mathematical, and have not yet been used to determine how the morphology of light curves due to moons depends on the system parameters.

Nevertheless, we can explore some general considerations that will help to elucidate how the detectability of moons depends on the system parameters, which in turn will help in the interpretation of our detailed simulations.  The size, shape, and orientation of the caustics created by moons depends on $q_m$, $s_m$, $q_p$, $s_p$, and $\Psi$.  For fixed values of the other parameters, $\Psi$ can affect the distance of the lunar caustic from the planetary caustic, as well as the size of the lunar caustic (see Appendix \ref{app:aligned}).  This may be somewhat counter-intuitive, as for an isolated planet+moon system, these quantities are independent of $\Psi$. For planets closer to the star, the gravitational shear due to the star breaks the symmetry.  Thus, all else being equal, as $s_p$ increases the dependence of the morphology and thus detectability of moons on $\Psi$ decreases.  

Gravitational shear is the effect of a lens object distorting an image created by another lens (see, e.g., \citet{Mao:92}). For example, an image created by a bound planet does not behave solely like a single lens; the lens star will have a significant impact on the characteristics of that image, the strength of which is given by the shear. For the moons we consider in our simulations, we first define
%Turning our attention to the other four parameters, it is helpful to consider three regimes. 
%We first define 
\begin{equation}
\gamma_s \equiv s_p^{-2}
\end{equation}
which is the dimensionless gravitational shear on the moon due to the star and 
\begin{equation}
\gamma_p \equiv s_m^{-2}
\end{equation}
which is the dimensionless gravitational shear on the moon due to the planet. Neither of these definitions involve the mass ratio ($q_p$ or $q_m$) because $s_p$ is normalized by $\theta_E$ (see Eq. \ref{eqn:thetae}), which depends on the total mass of the system (roughly the mass of the star), and $s_m$ is normalized by $\theta_{E,p}$, which depends on the mass of the planet. 
Note that the gravitational shear on the planet and moon due to the star will be of similar magnitude if $s_m q_p^{1/2}\ll s_p$.  With these definitions, we can consider three regimes.  
\begin{itemize}
    \item $\gamma_p\gg \gamma_s$ In this case, the shear on the moon is dominated by that from the planet.  This is the limit of an isolated (or free-floating) planet with a moon and is well approximated by a binary lens with parameters $q_m$ and $s_m$.   The size, shape, and location of the `planetary' caustics produced by the moon, e.g., the caustics produced when the moon perturbs one of the two images created by the planet, can be approximately represented by the analytic equations given in \citet{Han_2006} with angular sizes normalized to $\theta_{{\rm E},p}$.  This regime was explored by \citet{Chung:2016} in the context of moons orbiting wide-separation planets and \citet{Sajadian:2023} in the context of moons of free-floating planets. 
    \item $\gamma_p \ll \gamma_s$ In this case, the shear on the moon is dominated by the shear from the star.  In this regime, the perturbations due to the planet and moon can be approximated by their superposition \citep{Han:2001,Han:2008}, e.g., the magnification can be approximated as the superposition of two binaries, the planet-star binary with mass ratio $q_p$ and $s_p$ and the moon-star binary with mass ratio $q_p q_s$ and separation $s_p-q_m s_m\cos\Psi$.  
    \item $\gamma_s \sim \gamma_p$ In this case, the shear on the moon due to the star is comparable to the shear on the moon due to the planet.  This is the most phenomenologically rich regime, as the size, shape, and location of the caustics produced by the moon and planet depend on all five parameters $q_p, s_p, q_m, s_m$, and $\Psi$.  In particular, the shear due to the star and planet can combine constructively or destructively to increase or decrease the size of the caustic produced by the moon.  See Appendix \ref{app:aligned} for more discussion. 
\end{itemize}      
From these considerations, we can make the following arguments.  In the case of $\gamma_p\ll \gamma_s$, the conditional probability of detecting a moon given the detection of its host planet and the star $P({\rm moon|planet,star})$ is roughly the detection probability of the planet given the star in the absence of the moon times the detection probability of the moon given the planet in the absence of the star, or $P({\rm moon|planet,star})\simeq P({\rm planet|star})P({\rm moon|planet)}$. This assumes that the moon neither suppresses or enhances the planet detection probability.  In the opposite case of $\gamma_p\gg \gamma_s$, $P({\rm moon|planet,star})$ is just the probability of detecting the perturbation of the planet on the host star light curve in the absence of the moon times the probability of detecting the perturbation of the moon on the host star light curve in the absence of the planet, or $P({\rm moon|planet,star})\simeq P({\rm planet|star})P({\rm moon|star)}$. This assumes that caustics binary superposition approximation holds, which will not be true when $s_m \la 1$.  It also assumes that the moon can't be detected through the distortion it makes on the planetary caustic.  As we discuss in Section \ref{subsec:channels}, some moons may be detectable through this channel.  In the intermediate regime, the moon detection probability becomes much more complicated, because the presence of the planet can enhance or suppress the signature of the moon, and thus $P({\rm moon|planet,star})$ cannot be factored.

Given that moons are only stable if they orbit within $\sim 0.5 a_{\rm Hill}$ for prograde satellites and $\sim a_{\rm Hill}$ for retrograde satellites, and that the detection probability of the moon is generally highest for moons with $s_m \sim 1$ or semimajor axes near the Einstein ring radius of the planet $R_{{\rm E},p}=D_l\theta_{{\rm E},p}$, it is useful to consider the ratio between the Hill radius of the planet to its Einstein ring radius
\begin{equation}\label{eqn:hilltoRE}
s_{\rm Hill} \equiv \frac{a_{\rm Hill}}{R_{{\rm E},p}}=3^{-1/3} s_p q_p^{-1/6}.
\end{equation}
Note that $s_{\rm Hill}$ is weakly dependent on the planet/star mass ratio $q_p$ since the increase in the planetary Einstein ring radius with $q_p$ ($R_{{\rm E},p} \propto q^{1/2}$) is nearly exactly offset by the increase in the Hill radius with $q_p$ ($a_{\rm Hill} \propto q_p^{1/3}$). The net result is that this ratio decreases with increasing planet/star mass ratio as $s_{\rm Hill} \propto q_p^{-1/6}$. Thus the optimal region of parameter space for the detection of moons via microlensing depends primarily on the semimajor axis of the planet, and to a much lesser extent on its mass.

Since we are interested in the {\it conditional} probability that a moon is detected given the detection of the planet, it is also important to consider the detectability of the planet itself.  As is well known (e.g., \citealt{Gould_1992,Griest:1998,Bennett:1996}) the detection probability of exoplanets with microlensing increases for more massive planets and is maximized for planets with $s_p\sim 1$.  To provide a rough guide to the detection probability of a planet, we assume the cross-section for detection is proportional to the cross-section to encounter the planetary caustic(s).

In the close caustic regime ($s_p < s_c$), there are two symmetric triangular planetary caustics centered a distance of $\sim 1/s_p-s_p$ from the star, with one above the planet-star axis and the other equidistant from the planet-star axis but below it. We estimate the cross section to the planetary caustic in units of the stellar Einstein ring radius, $s_{caus}$, to be the geometric mean of the horizontal width of each caustic ($\Delta \eta$) and the vertical distance between the two caustics plus the sum of their vertical widths ($\Delta \xi$). See Figure \ref{fig:geometry} for visualizations of $\Delta\eta$ and $\Delta\xi$. We adopt the analytic estimates for the caustic sizes and positions from \citet{Han_2006}, 
\begin{equation}\label{eq:close_caustic_width}
    \Delta \eta = \frac{4q_p^{1/2}}{s_p(1-s_p)^{1/2}},\;\;\;  \Delta \xi = \frac{3^{3/2}}{4} q_p^{1/2} s_p^3,
\end{equation}
which gives
\begin{eqnarray}
   s_{caus} = \sqrt{\Delta \xi\Delta \eta } 
     &=& 3^{3/4}q_p^{1/2}s_p(1-s_p)^{-1/4} \\ 
     &\rightarrow& 2.28 q_p^{1/2}s_p, \label{eq:s_caustic_close}
\end{eqnarray}
where the last limit holds for $s_p\ll 1$.

In the wide caustic regime ($s_p > s_w$), there is one astroid-shaped caustic centered at a distance of $\sim s_p-1/s_p$ from the star. We again estimate $s_{caus}$ to be the geometric mean of the horizontal ($\Delta \xi$) and vertical width ($\Delta \eta$) of this caustic. We have that
\begin{equation}\label{eq:wide_caustic_width}
    \Delta \xi = \frac{4 q_p^{1/2}}{s_p (s_p^2-1)^{1/2}},\;\;\;    \Delta \eta = \frac{4 q_p^{1/2}}{s_p (s_p^2+1)^{1/2}},
\end{equation}
which gives,
\begin{eqnarray}
   s_{caus} = \sqrt{\Delta \eta \Delta \xi} &=& 4 q_p^{1/2}s_p^{-1} (s_p^4-1)^{-1/4}\\
     &\rightarrow& 4 q_p^{1/2}s_p^{-2},\label{eq:s_caustic_wide}
\end{eqnarray}
where the last limit holds for $s_p\gg 1$.

Considering the dependence of $s_{\rm Hill}$ and $s_{caus}$ on $q_p$ and $s_p$, we can make the following general observations.  First, since $s_{\rm Hill} \propto s_p q_p^{-1/6}$, we expect the probability of detecting a stable moon to depend weakly on $q_p$ but increase with increasing $s_p$, as moons orbiting wider planets that have larger Hill radii are less likely to have moons with $s_m \ll 1$, which generally produce weaker perturbations.  However, the detection probability of planets decreases with increasing $s_p$ as $\propto s_p^{-1}$ for $s_p\gg 1$.  Thus, we expect the conditional probability of detecting a stable moon given the detection of its host to peak at a value of $s_p \ge 1$, but the location of this peak should depend weakly on $q_p$.

\begin{figure*}
    \centering
    \includegraphics[width=0.99\textwidth]{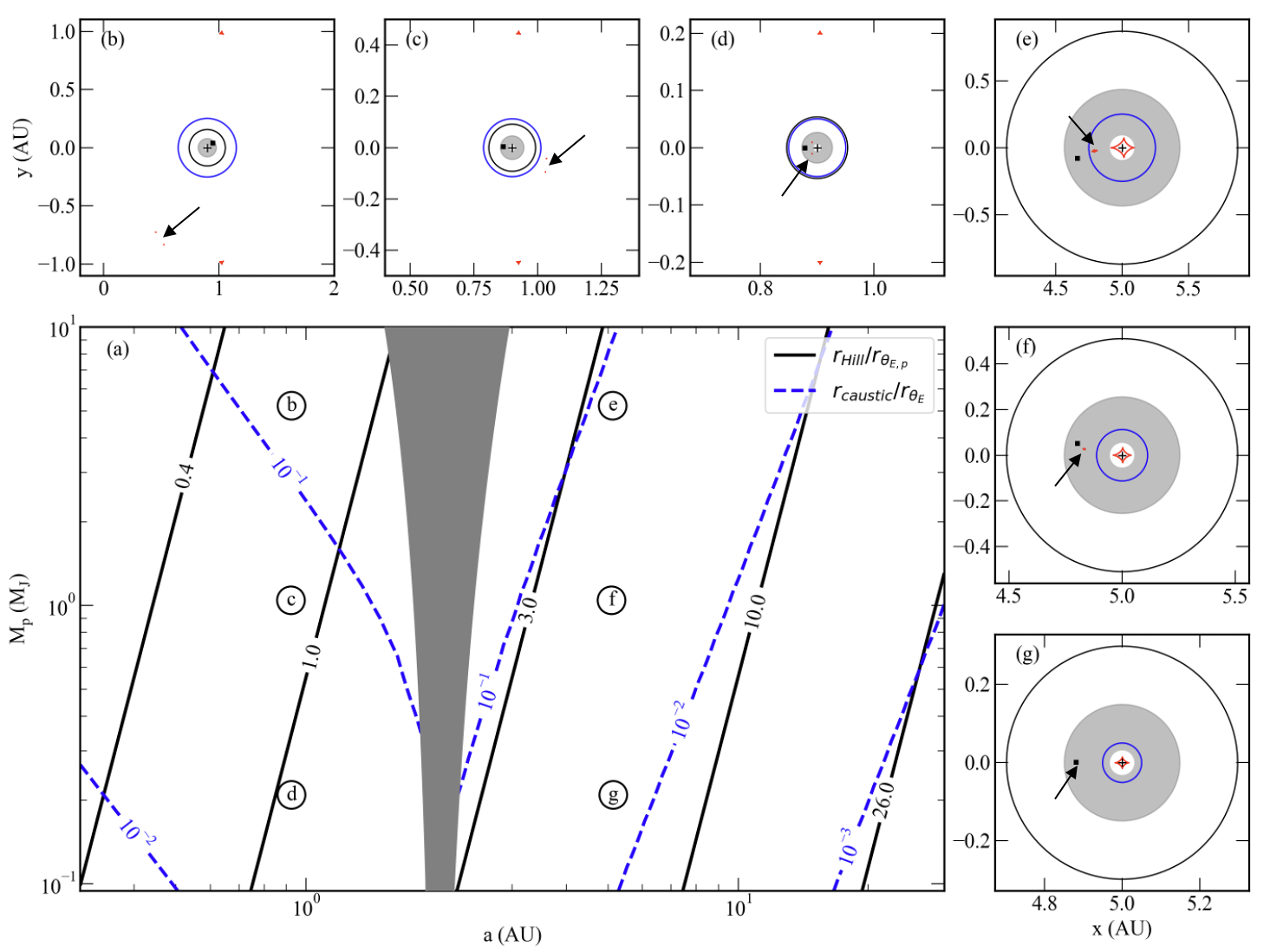}
    \caption{Panel a shows the range of semimajor axes $a_p$ and planet masses $M_p$ for which we simulated companion moons.  We assume a $0.3\ M_{\odot}$ host star, which is approximately the mean mass of the stars in our simulation. We assume a distance to the lens, $D_l$ of 7.37 kpc, which is approximately the front of the Galactic Bulge.
    The black solid lines show contours of constant $s_{\rm Hill} \equiv a_{Hill} / R_{{\rm E},p}$, e.g., the ratio of the Hill radius of the planet to the Einstein ring radius of a planet. The dashed blue curves show contours of constant $s_{caus}$, the mean caustic cross section (See Equations \ref{eq:s_caustic_close} and \ref{eq:s_caustic_wide}). Note that the definition of $s_{caus}$ for the close and wide regimes are slightly different. For the wide regime, $s_{caus}$ is simply related to the size of the planetary caustic. For the close regime, $s_{caus}$ also takes into account the vertical separation between the two planetary caustics. The gray, shaded region shows the region of the resonant caustic topology that we do not include in our simulations. Each offset panel (b-g) shows the caustic structures (in red) the Hill radius (in black) and the planetary Einstein ring radius (in blue) centered at the location of the planet (the plus symbol). The caustics were calculated using the {\tt triplelens} package \citep{Kuang_2021}. 
    The gray, shaded region in the inset is the region around the planet where we simulated moons (0.1-0.5 Hill radii). A moon with a randomly chosen semimajor axis and orbital phase is shown as a square. The arrows in each panel point to the locations of the small caustics from the moon. These panels are connected to the corresponding planet mass and semimajor axis in the main panel. This figure is discussed in more detail at the end of Section \ref{sec:general}.}
    \label{fig:caustic_hill}
\end{figure*}

Figure \ref{fig:caustic_hill} graphically illustrates the previous points. The main panel (a) shows the region of planet mass versus semimajor axis parameter space we are simulating. For this plot, we have assumed a stellar mass of 0.3 $M_{\odot}$, which is close to the average mass of the stars in our simulation. 

The gray-shaded region is the resonant caustic region, which we do not consider. At smaller semimajor axes the planetary perturbations are in the close caustic regime ($s_p < s_c$), while at larger semimajor axes the planetary perturbations are in the wide caustic regime ($s_p > s_w$).  In both the close and wide regimes, the size of the caustic increases for increasing planet mass. This means that more massive planets are easier to detect. In the close regime, the size of the caustic increases for increasing $s_p$, while in the wide regime, the size increases for decreasing $s_p$. Essentially, the size of the caustic peaks near $s_p=1$. Thus planets are most detectable when they are located near the stellar Einstein Ring radius \citep{Gould_1992,Griest:1998,Bennett:1996}.

The black lines show contours of constant $s_{\rm Hill}$, the ratio of the Hill radius to the planetary Einstein ring radius. This ratio increases for increasing semimajor axis, since for a constant planet mass, $s_{\rm Hill} \propto s_p$, and decreases with increasing planet mass, since for a constant semimajor axis, $s_{\rm Hill} \propto q_p^{-1/6}$. The blue-dashed lines show contours of constant $s_{caus}$.

Figure \ref{fig:caustic_hill} also contains six offset panels (b-g) that show examples of planet-moon systems from our simulations. The axes of these plots are in units of AU.  These panels are connected to a plus symbol in the main panel, which marks the mass and semimajor axis of the planet in the offset panel.  The Hill radius of the planet is shown in black and the planetary Einstein Ring radius is shown in blue. The gray-shaded region shows the region of the Hill radius that we populate with moons (0.1-0.5 $a_{\rm Hill}$). The black dot shows a moon with a randomly chosen semimajor axis in this range and a randomly chosen orbital phase $\alpha$ (see Section \ref{sec:romansim} for how the phase $\alpha$ relates to the projected angle $\Psi$).  The red curves show the caustics for the planet and moon, with the center of the planetary caustic shifted to the location of the planet.  We note that, in reality, the caustic is located at a distance of $|s_p-1/s_p|$ from the star. 

For the three close caustic cases (panels b-d), the planetary Einstein Ring radius is similar in size or larger than the Hill radius, making the simulated region well within the Einstein Ring, and making the moons more difficult to detect. For the wide caustic regime (panels e-g), the Hill radius is much larger than the planetary Einstein Ring radius, such that the Einstein Ring radius is located within the simulated region. Therefore, moons around these wide separation cases are much easier to detect, and we expect our simulations to detect a higher number of moons orbiting planets in the wide caustic regime rather than the close caustic regime.

\section{Assumed Distribution of Planet and Moon Properties}
\label{sec:distribution}

Here we focus on the detectability of giant moons orbiting giant planets.  Giant planets will be the easiest planets to detect with Roman \citep{Penny_2019}.  Massive moons orbiting these planets are likely to form in circumplanetary disks and may be a common by-product of planet formation (e.g., \citealt{Canup_2002}).  We do not consider systems analogous to our own Earth-Moon system for two reasons. First, the Moon almost certainly formed in a manner very different than the other giant moons in our Solar System, i.e., via a giant impact \citep[see, e.g.,][]{Canup:2004}.  The distribution of masses and orbits of moons formed by giant impacts is likely to be very different from those formed in a circumplanetary disk, as is their frequency.  Second, planets with masses similar to the Earth have very different demographics and detectability than the giant planets we consider here.  For example, the perturbations due to these types of systems are more likely to be impacted by strong finite source effects \citep{Han:2002}.  Nevertheless, exploring the detectability of Earth-Moon analogs with Roman is an interesting topic of future study. 

For the properties of the moons, we adopt a log-uniform distribution in $q_m$ from $10^{-4}$ to $10^{-2}$. Comparing to the giant moons in our Solar System, $10^{-4}$ corresponds roughly with the mass ratios of the Galilean moons of Jupiter, whereas $10^{-2}$ corresponds to approximately the Moon-Earth mass ratio.   We choose to adopt a distribution that is log-uniform in $q_m$ rather than $M_m$ as previous studies have suggested that, for moon formation in circumplanetary disks, there is a common mass scaling such that the total mass in moons is $\sim 10^{-4}-10^{-3} M_p$, where the coefficient depends on the detailed properties of the circumplanetary disk \citep{Canup_2006}.  The upper end of the range of $q_m$ we choose may therefore require an unusually massive circumplanetary disk.  

Finally, we choose the semimajor axis of the moon $a_m$ from a uniform distribution from $0.1$ to $0.5$ planetary Hill radii.  Moons orbiting within $0.1 a_{\rm Hill}$  will have values of $s_m \ll 1$, i.e., much smaller than the Einstein ring radius of the planet.  Such moons are unlikely to be detected, in analogy to the fact that planets with $s_p \ll 1$ are unlikely to be detected in planetary microlensing events \citep{Gould_1992,Liebig:2010}.  We choose a maximum of $0.5$ Hill radii as this corresponds to the largest stable orbit for moons on prograde orbits \citep{Domingos:2006}.  Finally, we assume that both the planet and moon are on circular orbits and that the orbit of the moon is coplanar with the orbit of the planet.

Thus we simulate planets with masses drawn from a log-uniform distribution from $30 M_{\oplus}$ to $10 M_J$ and semimajor axes drawn from a log-uniform distribution from $0.3$ to $30$ AU. We populate these planets with moons with mass ratios $q_m$ in the range $10^{-2}-10^{-4}$ and semimajor axes in the range $[0.1-0.5]a_{\rm Hill}$. We normalize the distribution function of planet masses and semimajor axes according to the following fiducial mass function adopted by \citet{Penny_2019}
\begin{equation}
    \frac{d^2 N}{d \log M_p d \log a} = 0.24 \; \textrm{dex}^{-2}\left( \frac{M_p}{95M_{\oplus}}\right)^{-0.73}
    \label{fiducialmf}
\end{equation}
This mass function is based on the power-law mass function of bound planets found by microlensing as determined by \citet{Cassan_2012}.  We note that the mass function of \citet{Penny_2019} saturates at a value of 2 planets per dex$^2$ per star below a mass of $5.2M_\oplus$.  However, since none of the planets we assume have masses less than $5.2~M_\oplus$, this saturation has no impact on our results. We normalize the distribution of moon mass ratios and semimajor axes such that there is one moon per planet over the range simulated. We note that, because we assume that every planet hosts a moon, but adopt the above planet mass function that increases toward lower planet masses, moons orbiting lower-mass planets are intrinsically more common in our simulations. 

\begin{figure}
    \centering
    \includegraphics[width=\linewidth]{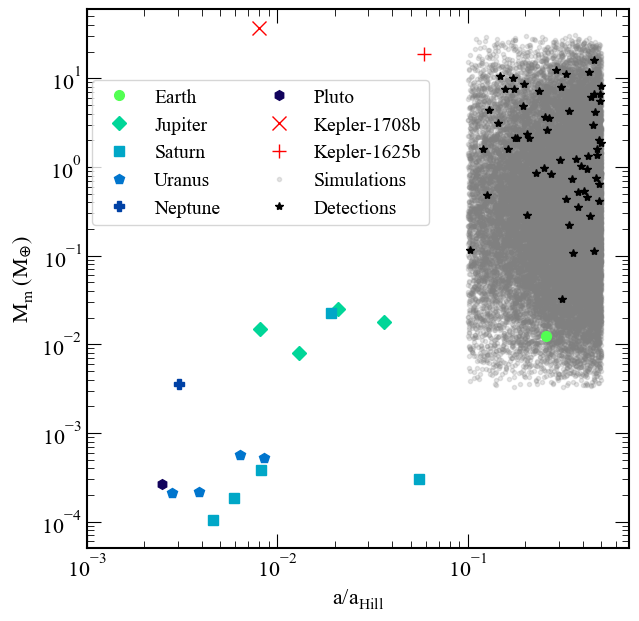}
    \caption{The masses of large Solar System moons in Earth masses versus their semimajor axes in units of the Hill radius. Each different symbol and color represents the moons of a different planet. The red symbols are the two exomoon candidates, Kepler-1708bi \citep{Kipping_2022} and Kepler-1625bi \citep{Teachey_2018}. The gray points are our simulated moons. Each of the black stars is a detected moon in our simulation.}
    \label{fig:ss_moons}
\end{figure}

Figure \ref{fig:ss_moons} shows the masses of the simulated moons in units of $M_{\oplus}$ as a function of their semimajor axes scaled to $a_{\rm Hill}$.  We compare these to massive Solar System moons and two exomoon candidates, Kepler-1708bi \citep{Kipping_2022} and Kepler-1625bi \citep{Teachey_2018}. None of the moons we simulate are directly analogous to the known moons in our Solar System in this parameter space except for our own Moon.   In particular, all of the giant moons orbiting the giant planets in our Solar System have semimajor axes within $0.1$ Hill radii of their host planet, which is generally well within the Einstein ring radius even for the most distant planets we consider (see Figure \ref{fig:caustic_hill}).  Although our Moon does fall into the region of $M_m$ and $a/a_{\rm Hill}$ parameter space we simulate, our results likely bring little to bear on the ability of Roman to explore the frequency of Moons like our own because the Earth is much less massive than $30~M_\oplus$, the smallest planetary mass we consider.  However, our simulation results are still of interest as it is not necessarily the case that the moons in the Solar System are representative of moons orbiting exoplanets, just as the Solar System planets are not generally representative of the known population of exoplanets.

\section{Roman Simulations} \label{sec:romansim}

For this work, we perform our simulations in a two-step process. The first step uses the {\tt GULLS} microlensing simulation tool to simulate large numbers of planetary microlensing events \citep{Penny_2019}. The {\tt GULLS} simulation tool provides realistic distributions of stellar host properties (e.g., $M$, $D_l$, AB magnitude) and microlensing event parameters (e.g., $\theta_{\rm E}$, $\mu_{\rm rel}$, $t_{\rm E}$), as well as simulated data of the microlensing event with the photometric properties expected from Roman.  

The photometric noise model we adopt here is shown in Figure 4 of \citet{Penny_2019}  This noise model is similar to, but somewhat more conservative than, the more recent noise model used in \cite{Wilson:2023}, which is shown in their Figure 4 (see also their Figure 9).  The true noise per epoch will differ from what we have assumed both due to changes in the final survey design (in particular cadence and exposure time) relative to what was assumed by \citep{Penny_2019} (see the recently released ROTAC report\footnote{https://roman.gsfc.nasa.gov/science/ccs/ROTAC-Report-20250424-v1.pdf}), and due to the on-orbit performance of the telescope, instrument, and pipeline will differ from simulations.  Given the relatively low yield of exmoons that we predict we expect that the Poisson fluctuations in this yield will be larger than differences in the yield we would predict using a more realistic noise model. 

We use the same Galactic Model as Penny et al. 2019 in our simulations, namely version 1106 of the Besancon Galactic model (\cite{Robin2003, Robin2012}, see \cite{Penny_2019}). We only provide a brief overview here, and refer the reader to \cite{Penny_2019} for the full details. 

{\tt GULLS} draws source and lens stars from star catalogs produced by a Galactic model and uses them to simulate individual microlensing events. Each event is given a normalized weight, $w_i$, which is proportional to that event's contribution to the total event rate along that line of sight. The sum of these weights gives the number of microlensing events expected along that line of sight during the duration of the survey. A planet is added to each event according to the distribution outlined in Section \ref{sec:distribution} and the corresponding binary lens light curve is calculated. A single lens model is then fit to the simulated binary lens, and the values of $\chi^2$ for the binary lens ($\chi^2_p$) and the single lens ($\chi^2_{s}$) are determined, as well as the $\Delta \chi^2_p=\chi^2_p-\chi^2_s$ between the binary lens and single lens fits.  The planet is considered detected if $\Delta \chi^2_p > \Delta \chi^2_{thr,p}=160$ \citep{Penny_2019}. The result is a catalog of events with detected planets.

One important feature of {\tt GULLS} is that it provides realistic simulated light curves with the expected photometric properties from Roman including contributions to all noise sources (source flux, lens flux, blend flux, read noise, background, etc.). The net result is a more accurate determination of the sensitivity of Roman to exomoons. 

To simplify our calculations, we only simulate events in a single square 0.25x0.25deg$^2$ area centered at $l=1.1^\circ$, $b=-1.7^\circ$ over a single observing season.  However, we assume a cadence appropriate for 7 fields assuming 48.6 s total exposure times and slew and settle times from the Cycle 7 Roman design, as was done in \citet{Penny_2019}.  We also assume that there will be six seasons, each with 72 days. We compute the nominal event rate over the entire Roman footprint in 0.25x0.25deg$^2$ tiles using the Besancon Galactic Model v1106 (\cite{Robin2003, Robin2012}, see \cite{Penny_2019}),  and then find the average event rate weighted by the area each of these tiles that is covered by the Roman footprint. Thus, to obtain yields for the entire survey, we assume even sensitivity across all fields and seasons, and then multiply our single-season, single-tile yields by the total area Roman covers (1.97 deg$^2$) and the number of seasons (6).  These total yields can be compared directly to the yields computed by \citet{Penny_2019}, with the only difference being the second-order effect of the varying yield across the Roman footprint of the survey due to the changing microlensing event rate and the dependence of microlensing event properties on Galactic coordinates. 

The second step of our simulation injects moons into the light curves of detected planets.  We calculate the light curves for these systems in the region proximate to the planetary perturbation using the inverse ray-shooting algorithm with the remainder of the light curve computed using {\tt MulensModel} \citep{Poleski:2019}, as described in Section \ref{subsec:rayshoot}. We compute two light curves for each event, one that includes the moon and one that does not, using the event parameters generated by {\tt GULLS}.  We then fit both light curves computed in this way to a planet-only model computed using {\tt MulensModel}  \citep{Poleski:2019}. We then use the difference in $\chi^2$ of these two fits to determine the significance of the moon perturbation.  By calculating both planet and planet+moon perturbations using inverse ray shooting, but fitting both to an essentially noise-free model, we can partially calibrate the effect of statistical noise from our magnification map calculations on the estimated $\Delta\chi^2$.  As a result, this metric more directly reflects the significance of the perturbation due to the moon. All other elements of the event, including the source radius $\rho$, the source trajectories, the photometric uncertainties, etc. are fixed to the values from the {\tt GULLS} simulation.

The moons we inject into our simulations are chosen and defined by three parameters: $q_{m}$, the moon-planet mass ratio; $s_m$ the moon-planet separation in units of $\theta_{{\rm E},p} = q^{1/2} \theta_E$; and $\Psi$, the angle between the planet-star axis and the moon-planet axis.  As noted above, we draw $q_m$ from a log-uniform distribution from $10^{-4}$ 
to $10^{-2}$. We choose $a_m$ from a uniform distribution from $0.1$ to $0.5$ Hill radii.  We assume that the moon and planet orbits are coplanar, and draw an inclination $i$ for these orbits from a uniform distribution in $\cos{i}$ from $(0,1]$.  We then assume a random phase $\alpha$ of the moon's
orbit from $(0,2\pi]$.  Then the component of the moon's projected separation from the planet along the planet-star axis is $s_{m,x} = -(a_m/R_{{\rm E},p})\cos(\alpha)$ and perpendicular component is $s_{m,y} = (a_m/R_{{\rm E},p})\sin(\alpha)\cos(i)$.  Finally, $s_m=(s_{m,x}^2 + s_{m,y}^2)^{1/2}$ and  $\Psi = \arctan(s_{m,y}/s_{m,s})$.

For this work, we choose to exclude any case where the planet is in the resonant caustic topology.  We find that these events are too computationally expensive to be efficiently modeled with the inverse ray-shooting method outlined above. The effects of the exclusion of these events on our results are discussed in Section \ref{subsec:resonant events}. We also choose to exclude any planetary event due to a central caustic perturbation for the same reason.

To determine whether our simulated event includes a detected moon, we adopt a $\Delta\chi^2$ criterion. We fit each of our simulated light curves (with and without a moon) with a binary lens model using {\tt MulensModel} \citep{Poleski:2019} and the following free parameters: $q_p, s_p, u_0, t_0, t_E, \alpha$, $\rho$, $F_S$ and $F_B$ where $F_S$ is the source flux and $F_B$ is the flux of any unrelated stars that are blended with the source. We calculate the $\chi^2$ value for each fit and determine the difference between the two $\chi^2$ values
\begin{equation} \label{eq:detect thres}
    \Delta \chi^2_m = \chi^2_m - \chi^2_p 
\end{equation}
Here $\chi^2_m$ is for the binary lens fit to the star+planet+moon light curve and $\chi^2_p$ is for the fit to the star+planet light curve. The moon is detected if $\Delta \chi^2_m>\Delta\chi^2_{thr,m}$.  We choose the threshold to be $\Delta \chi^2_{thr,m} = 90$. While this is nominally lower than the threshold for a single planet detection from \cite{Penny_2019}, we chose this lower threshold because we only search for moons near detectable planetary perturbations, whereas \cite{Penny_2019} was considering the detection of planetary perturbations at any point in the light curve.  Thus there is a lower probability that a spurious signal will occur near the planetary perturbation than that a spurious signal will occur at any point in the light curve.  This allows us to be less conservative in the $\Delta \chi^2$ threshold we require for the detection of a moon. 

\section{Results} \label{sec:results}

\begin{figure*}
    \centering
    \includegraphics[width=0.82\textwidth]{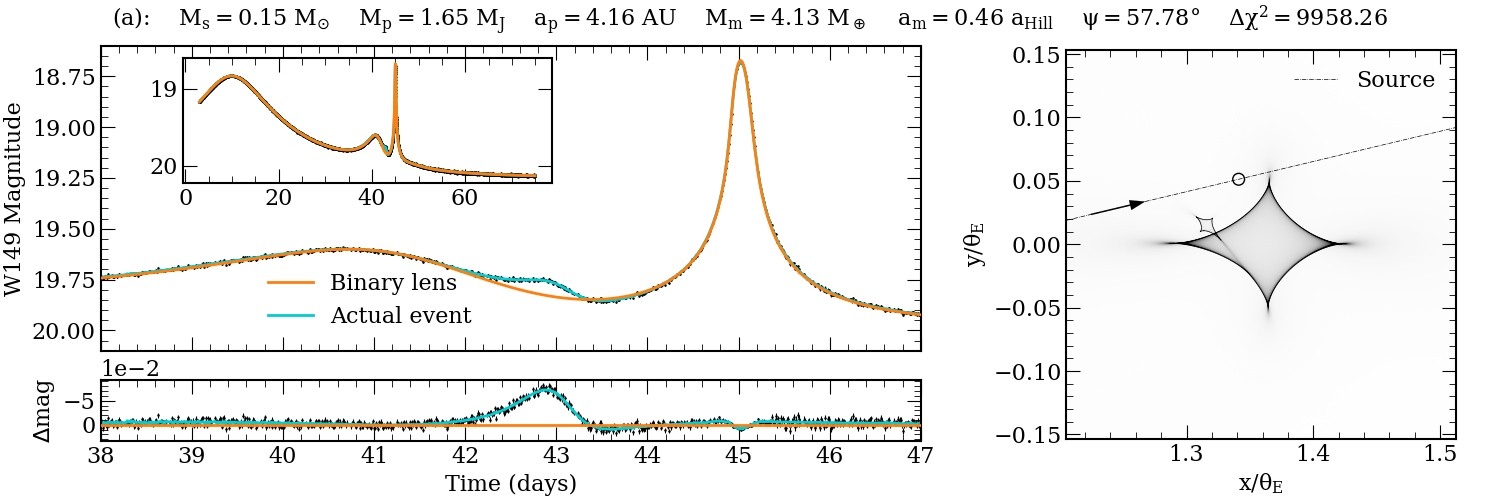}
    \includegraphics[width=0.82\textwidth]{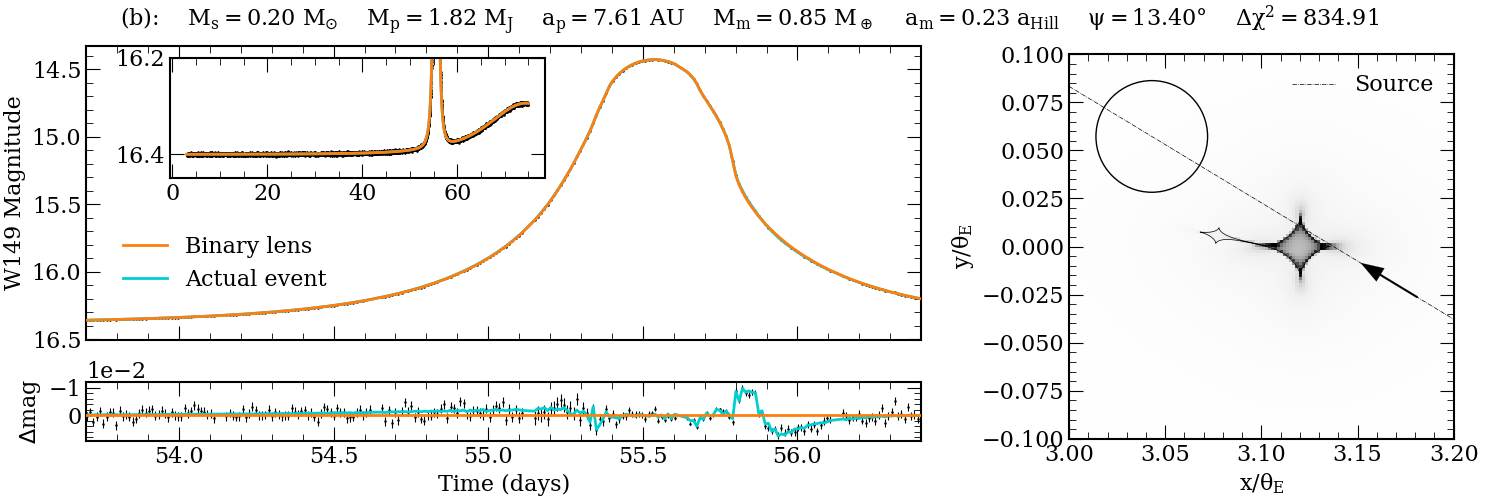}
    \includegraphics[width=0.82\textwidth]{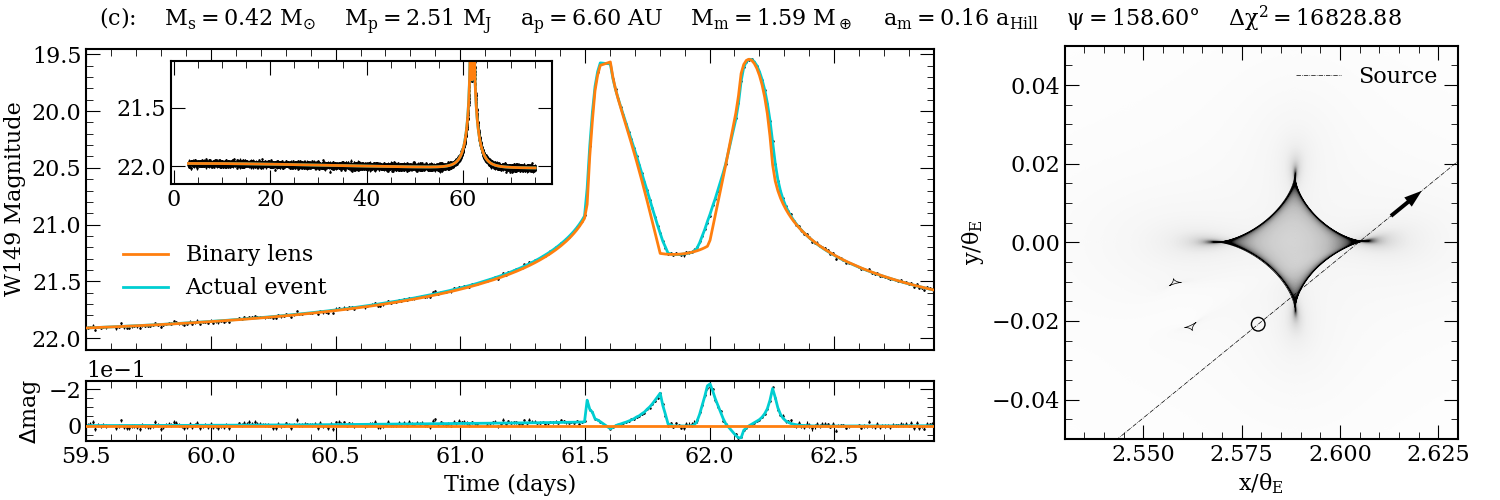}
    \includegraphics[width=0.82\textwidth]{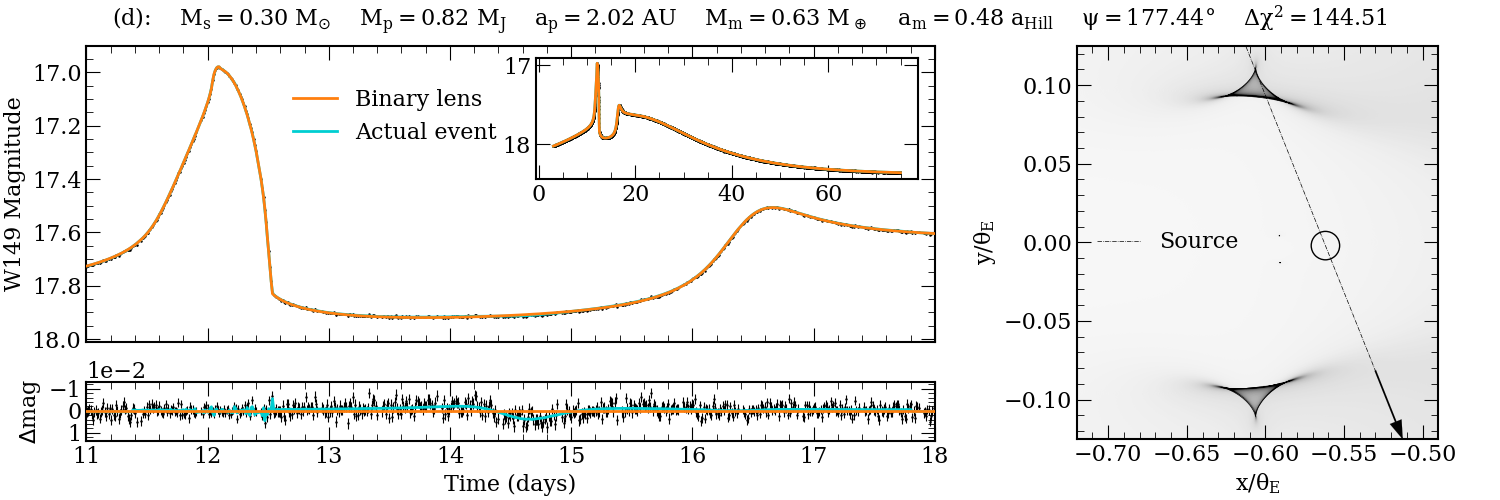}
\end{figure*}

\begin{figure*}
    \centering
    \includegraphics[width=0.81\textwidth]{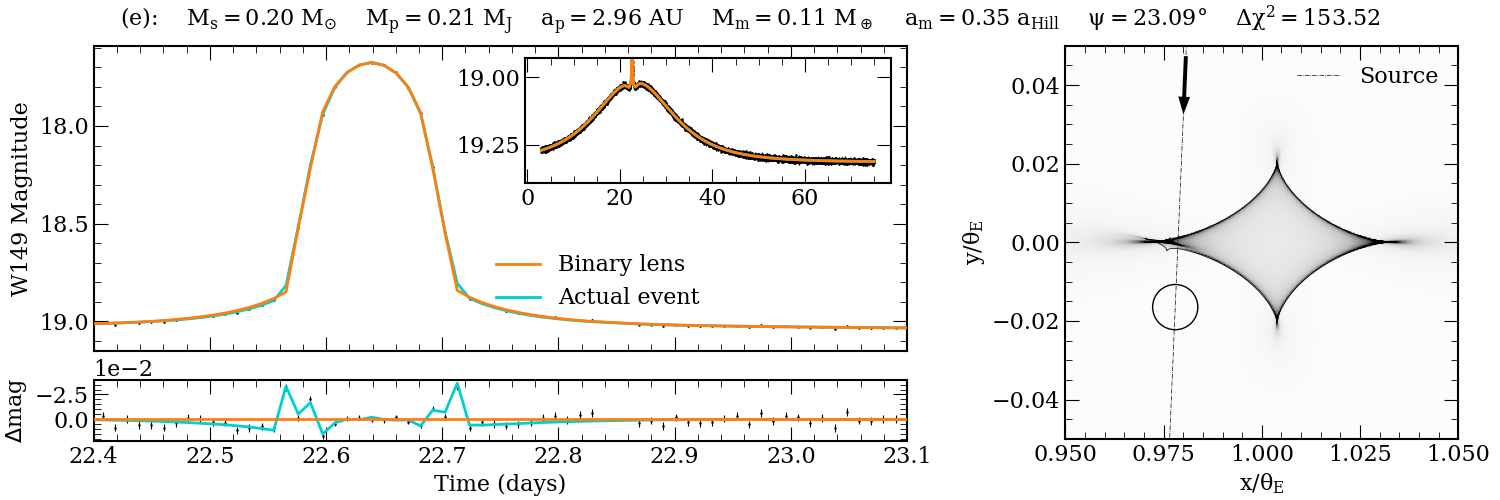}
    \includegraphics[width=0.81\textwidth]{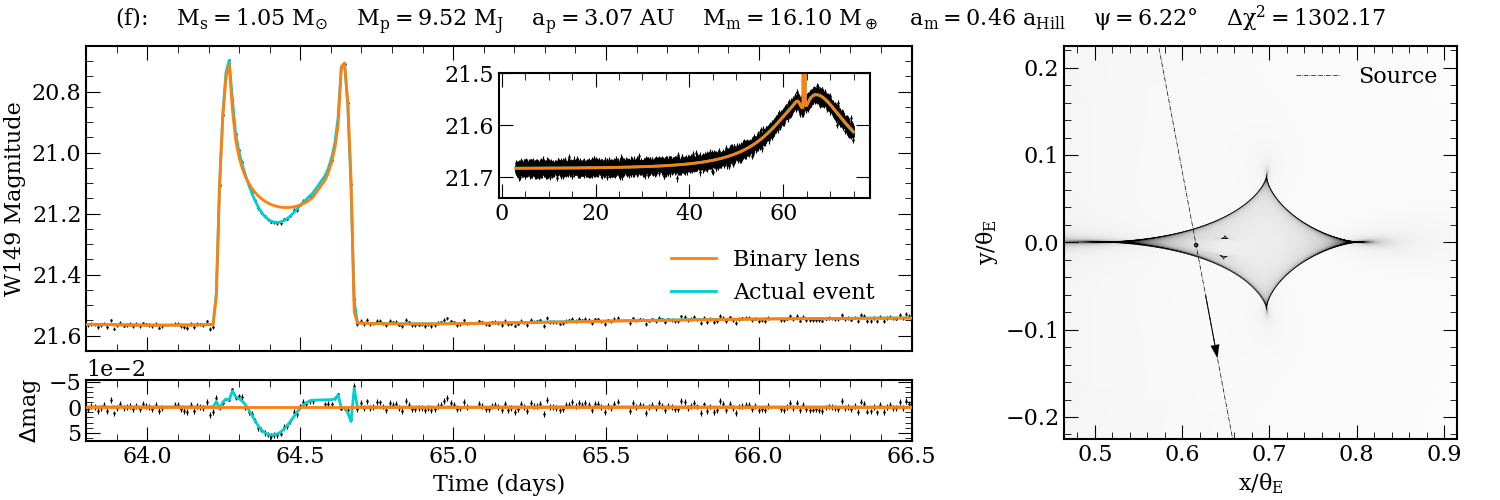}
    \includegraphics[width=0.81\textwidth]{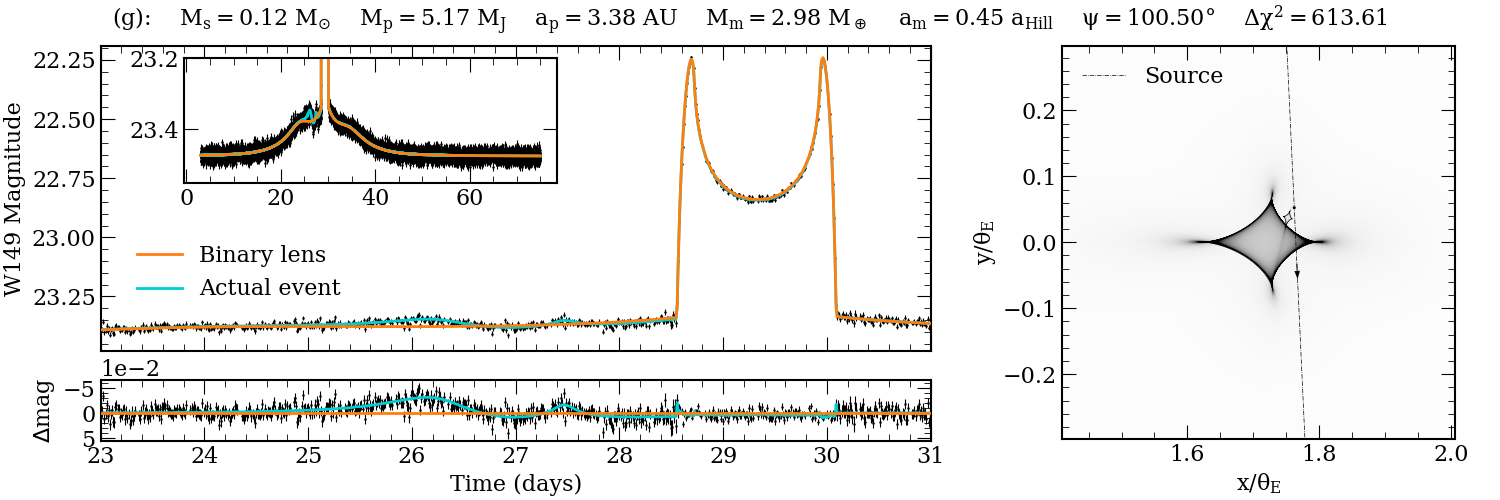}
    \includegraphics[width=0.81\textwidth]{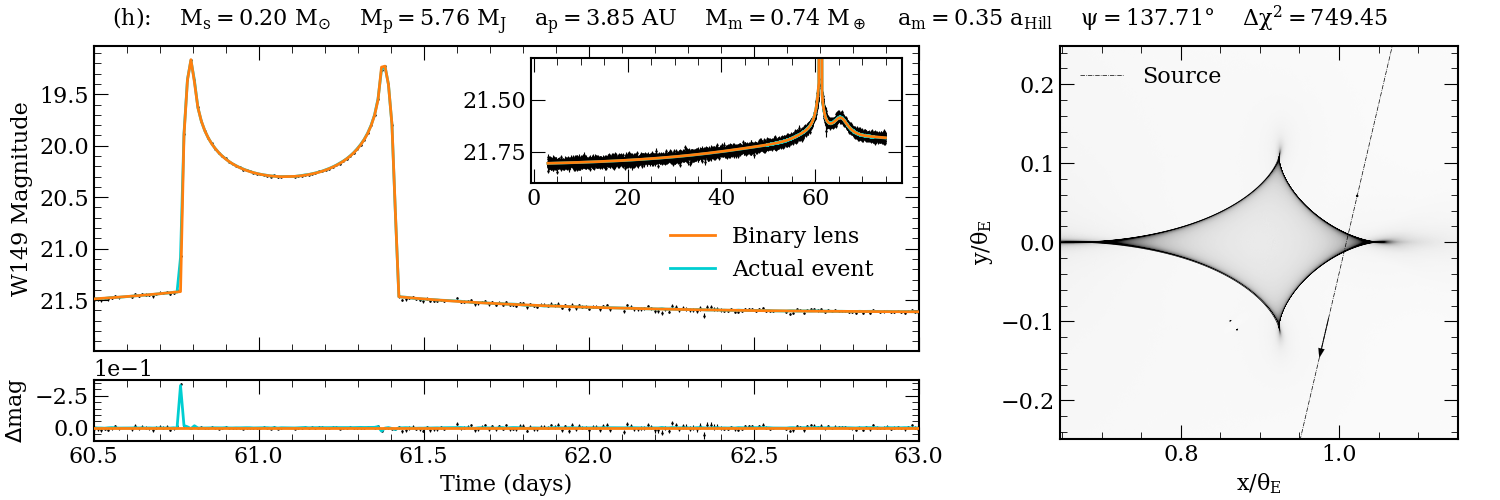}
    \caption{Examples of simulated Roman light curves with detectable moons. For each light curve, the left panel shows the region of the light curve near the planet and moon perturbations. The inset plot shows the full light curve. The black points with uncertainties are the simulated data points. The orange curve is the best fit binary lens model and the blue curve is the actual event including the moon. The bottom panel on the left shows the residuals of the planet+moon event from the binary lens fit.  The right panel shows the magnification map for this event. The dashed line shows the source trajectory, with the direction of the source given by the arrow. The circle along the source trajectory shows the size of the source. A variety of parameters for each event are listed above each plot. The events displayed here were chosen to display a variety of the features present from exomoons in our simulations.}
    \label{fig:lc b}
\end{figure*}

Here we present the results of our work and give a prediction for the number of exomoons Roman will detect given our assumptions. 

In total, we simulated $29,847$ events with detected planets and injected moons. Figure \ref{fig:lc b} shows a sample of the light curves of detected moons from our simulation.  These illustrate the diversity of moon perturbations we may expect to see with Roman.  Figures \ref{fig:lc b}b,e show examples of `buried' perturbations, where the source size is significantly larger than the moon caustic size.  This is analogous to the detections of `buried' planetary perturbations due to central and resonant caustics in high magnification events with strong finite source effects \citep{Dong:2009,Janczak:2010}.  Figures \ref{fig:lc b}a,b,g show events due to wide ($s_p>s_w$) planets and wide ($s_m \ga 1$) moons.  Figure \ref{fig:lc b}f shows an event due to a wide planet ($s_p>s_w$) and close ($s_m \la 1$) moon.  Figure \ref{fig:lc b}d shows an event due to a close ($s_p<s_c$) planet and close ($s_m \la 1$) moon; such detections are less common in our simulation.  Figure \ref{fig:lc b}c shows the example of a detection where the source does not pass directly over the caustic produced by the moon.  Rather, the moon is detected due to the fact that it distorts the planetary caustic in a way that cannot be reproduced by a binary lens.  See \citet{Gould:2014}. We discuss this class of detections in more detail in Section \ref{sec:discussion}.  Finally, Figure \ref{fig:lc b}h shows an example of the `detection' of a moon based on one highly deviant point.  The deviant point arises because the moon is distorting the planetary caustic, as with the detection in Figure  \ref{fig:lc b}c.   However, because the caustic crossing is short, the distortion only manifests as one deviant point.  We consider such events more carefully below.  

\begin{figure*}
    \includegraphics[width=0.99\textwidth]{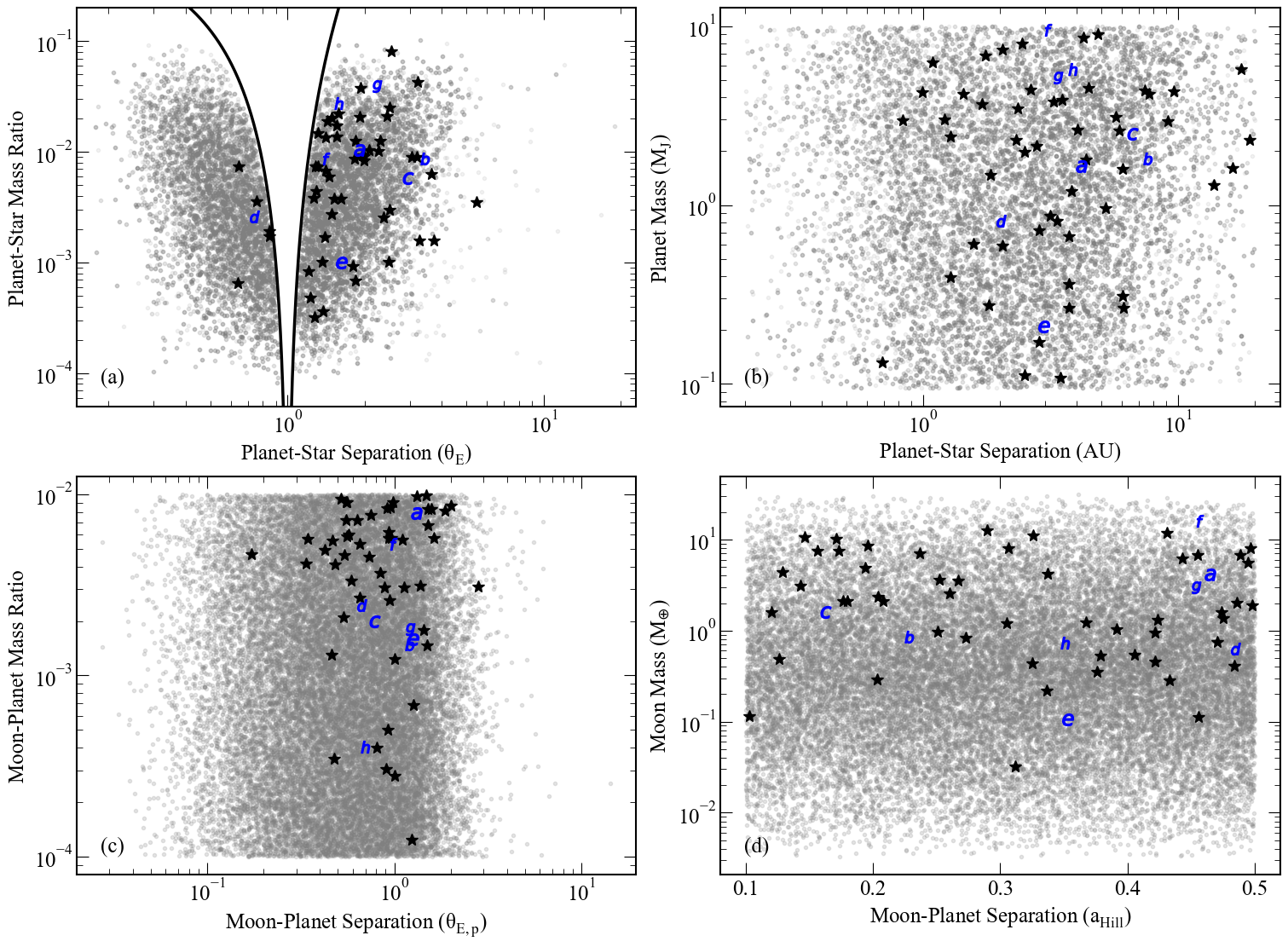}
    \caption{Physical parameters of the simulated events with and without detected moons. In each panel, the grey points show the events our simulation where a moon was not detected, i.e., these are the distributions of the physical parameters of the planets and moons we simulated. The stars show events where the moon was detected. The blue markers indicate the locations of the light curves shown in Figure \ref{fig:lc b} in these parameter spaces. Each panel shows (a) the planet/star mass ratio versus planet projected separation in units of the Einstein ring radius of the system. The lines indicate the boundaries of the resonant caustic region that we exclude from our sample; (b) the planet mass versus the planet semimajor axis; (c) moon/planet mass ratio versus the moon-planet separation in units of the Einstein ring radius of the planet; (d) moon mass versus semimajor axis of the moon in units of the Hill radius.}
    \label{fig:param plots}
\end{figure*}

\begin{figure*}
    \centering
    \includegraphics[width=0.9\textwidth]{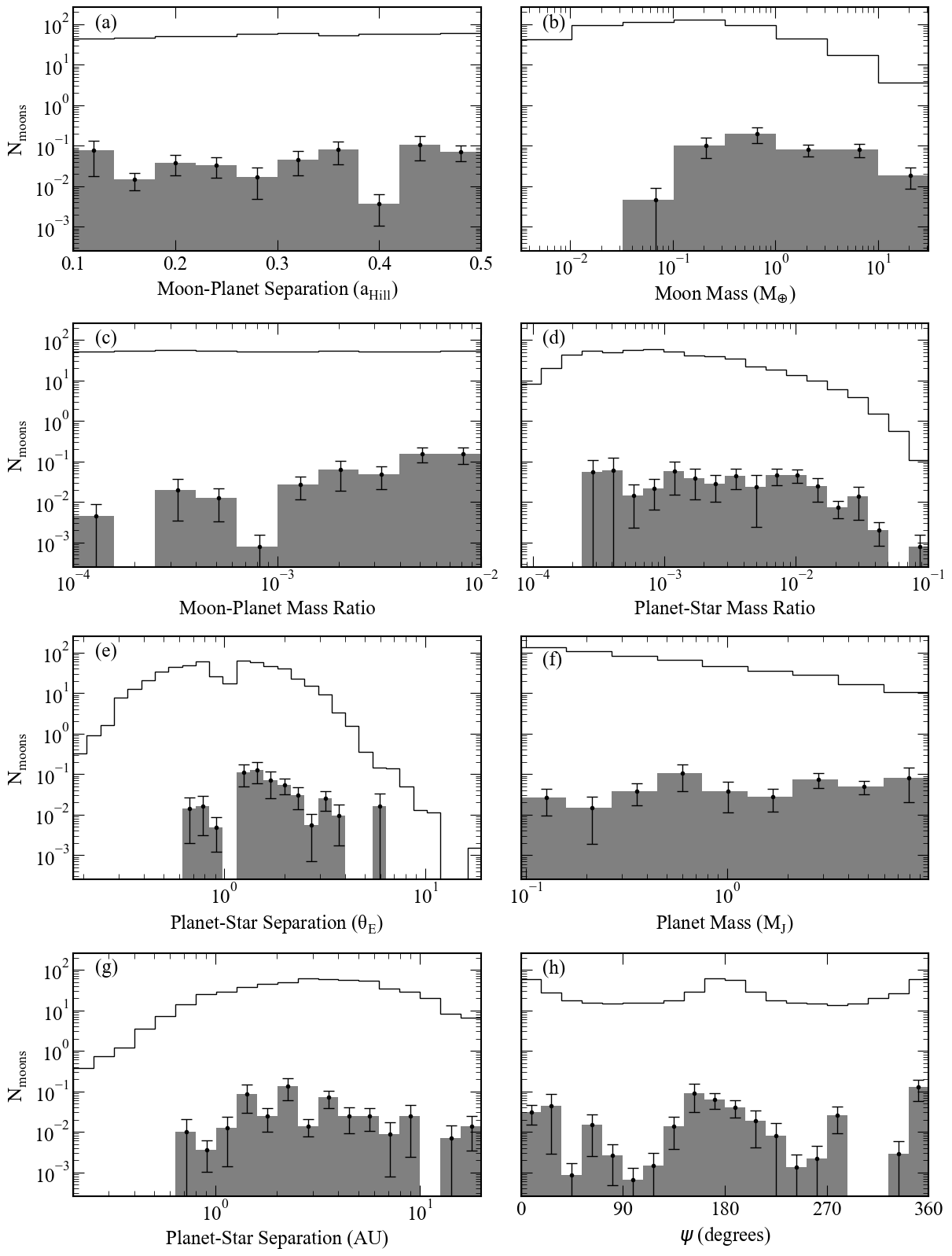}
    \caption{These histograms show the distribution of simulated and detected events. In each panel, the solid grey bars show the distribution of events with detectable exomoons. The solid line 
    shows the distribution of all simulated events. The height of each bin is the yield, or the number of events in each bin. The sum of the heights of all bins is the total predicted yield.}
    \label{fig:histograms}
\end{figure*}

Figures \ref{fig:param plots} and \ref{fig:histograms} show summary plots of the physical parameters of the simulated events with and without detectable moon perturbations.
Figure \ref{fig:param plots} plots every simulated event as a point, with the stars indicating those with exomoon detections and the grey points indicating non-detections (the blue letters denote the locations of the light curves from Figure \ref{fig:lc b}). Each panel shows the distribution of events and detections in different parameter spaces. 

Figure \ref{fig:param plots}(a) shows the parameters of both simulated and detected planetary events in natural microlensing units ($q_p$ versus $s_p$). The solid lines indicate the boundaries of the resonant caustic region, which is excluded from our sample. This region acts as a dividing line between close planetary caustic perturbations at smaller separations and wide planetary caustic perturbations at larger separations. We observe that most of the detected moons orbit planets biased toward wider separations compared to the simulated population. This bias arises from a competition: as the value of $s_{\rm Hill}$ increases with $s_p$, moons become easier to detect, but at the same time, the detection probability of the host planets decreases as $s_p$ increases.  Additionally, we find that most detected moons orbit planets with a planet-star mass ratio greater than $10^{-3}$. This is due to the competing factors of the simulated planet mass function, which favors lower-mass planets, the reduced detectability of planets with smaller mass ratios, and the smaller typical moon masses around lower-mass planets.  In this case, the latter two effects dominate over the first effect. 

Figure \ref{fig:param plots}(b) shows the planet mass versus the planet semimajor axis ($a_p$) for simulated events. We find that moons are generally detected around wider-orbit planets. The mean semimajor axis of planets that host detected moons is $\log{(a_p)} = 0.521 \pm 0.046$, while the mean semimajor axis for those without detections is $\log{(a_p)} = 0.374 \pm 0.002$. This is due to the increasing size of the Hill radius as the planet semimajor axis increases, allowing for wider orbit moons to exist.

Figure \ref{fig:param plots}(c) shows the moon-planet mass ratio $q_m$ versus moon-planet separation $s_m$. Most detected moons have a moon-planet mass ratio $q_m\ga 10^{-3}$ and $s_m\ga 0.5$. However, we do find detections with $q_m$ as low as $\sim 10^{-4}$ and $s_m$ as low as $\sim 0.2$.

Figure \ref{fig:param plots}(d) shows moon mass versus moon semimajor axis in units of the Hill radius for the simulated events. We find that exomoons are generally only detected when the mass of the moon is $\ga 0.2 M_\oplus$ or about twice the mass of Mars. We also find that Roman can detect exomoons at any separation between 0.1 and 0.5 times the planet's Hill radius. We discuss this somewhat surprising result below. 

\begin{table}
\centering
\caption{Exomoon yield as a function of moon mass, moon-planet separation in units of the Hill Radius, and planet mass. The width of each bin is 1 dex for the moon mass, 0.1 for the moon separation, and 0.5 dex for the planet mass, and the bins are centered at values listed in the table as the centers of each bin.  The bottom row shows the total number of detections.}
\label{tab:yields}
\begin{tabular}{>{\centering\arraybackslash}p{0.23\linewidth}>{\centering\arraybackslash}p{0.24\linewidth}|>{\centering\arraybackslash}p{0.13\linewidth}>{\centering\arraybackslash}p{0.24\linewidth}}
\hline
\textbf{Moon Mass ($M_{\oplus}$)} & \textbf{Yield} & \textbf{$a_m / a_{Hill}$} & \textbf{Yield} \\
\hline
$10^{-2}$ & $0.0 \pm 0.0$ & 0.15 & $0.11 \pm 0.06$ \\
$10^{-1}$ & $0.11 \pm 0.05$ & 0.25 & $0.07 \pm 0.03$ \\
1 & $0.28 \pm 0.09$ & 0.35 & $0.13 \pm 0.05$ \\
10 & $0.10 \pm 0.03$ & 0.45 & $0.18 \pm 0.07$ \\
\hline
Total: & $0.488 \pm 0.110$ & Total: & $0.488 \pm 0.110$ \\
\hline
\end{tabular}
\end{table}

Figure \ref{fig:histograms} shows histograms of the number of events simulated and the yield of detected moons detected as a function of several planet and moon parameters. The yield for each bin is calculated by summing the weights of every event in each bin and multiplying by the appropriate scale factors, as described in Section \ref{sec:romansim}. The parameter spaces are the same as in Figure \ref{fig:param plots}. We also include a histogram showing the number of events as a function of angle $\Psi$.  

There are a few trends worth highlighting. From Figure \ref{fig:param plots}c and Figure \ref{fig:histograms}a, we find that Roman's sensitivity to moons is nearly independent of $a_m/a_{\rm Hill}$. This is somewhat surprising, as we previously argued that we expect moons to be more readily detectable if they have semimajor axis such that $s_m \ga 1$.  This would seem to favor moons with larger orbits.  We find that the detected moons with smaller $a_m/a_{\rm Hill}$ tend to orbit planets with larger $a_p$.  Specifically, we find that the average semimajor axis of the planets hosting detected moons with $a_m/a_{\rm Hill}<0.3$ is $\langle a_p\rangle=6.3 \rm AU$, whereas $\langle a_p\rangle=3.2 \rm AU$ for planets hosting detected moons $a_m/a_{\rm Hill}>0.3$.  

We find that most detected moons have masses $M_m \ga 0.1M_{\oplus}$ and mass ratios $q_m\ga 3\times 10^{-4} $ (Fig.~\ref{fig:histograms}b,c)  Less massive and smaller moon/planet mass ratio moons create smaller caustics, which are less likely to be detected and tend to suffer from stronger finite source effects, which tend to suppress the perturbations.

Figure \ref{fig:histograms}e demonstrates that most detected moons orbit planets with $s_p \ga 1$, with a significant number orbiting planets with $s_p \ga 2$. The valley at $s_p \sim 1$ is a result of our exclusion of systems with resonant planetary caustics.  Planets with $s_p>2$ are outside of the traditional `lensing zone' ($s_p=0.6-1.6$) where the probability of detecting the planet is highest\citep{Gould_1992,Griest:1998}. The fact that a significant number of number moons are detected orbiting such planets even though the planets are themselves less likely be detected is because the planets have larger $a_{\rm Hill}$. We find that Roman will detect more moons orbiting planets with wide separations than close separations.  Similarly, Figure \ref{fig:histograms}g shows that moons are detectable around planets orbiting at a distance greater than $\sim 1$ AU from the star, with a peak for planets with semimajor axes $\sim 5$ AU.

Because we are assuming that the moon orbits are coplanar with the planet orbits, and because $\cos i$ is uniformly distributed such that edge-on systems are more likely, the intrinsic distribution of $\Psi$ for simulated moons peaks at $\Psi=0$ and $180^\circ$, and more moons have moon-planet separation vectors that are aligned and anti-aligned with the planet-star axis than perpendicular to the planet star axis. We find that the distribution of detected moon angles is even more clustered around $\Psi=0$ and $180^\circ$ than the input distribution.  We hypothesize that this is due to constructive (when $\Psi =0,180^\circ$) and destructive (when $\Psi = 180^\circ,270^\circ$ interference between the shear on the moon due to the planet $\gamma_p$ and the shear on the moon due to the star $\gamma_s$ when $\gamma_p \sim \gamma_s$. We explore this effect in more detail in the Appendix \ref{app:aligned}.

\begin{figure}
    \centering
    \includegraphics[width=0.47\textwidth]{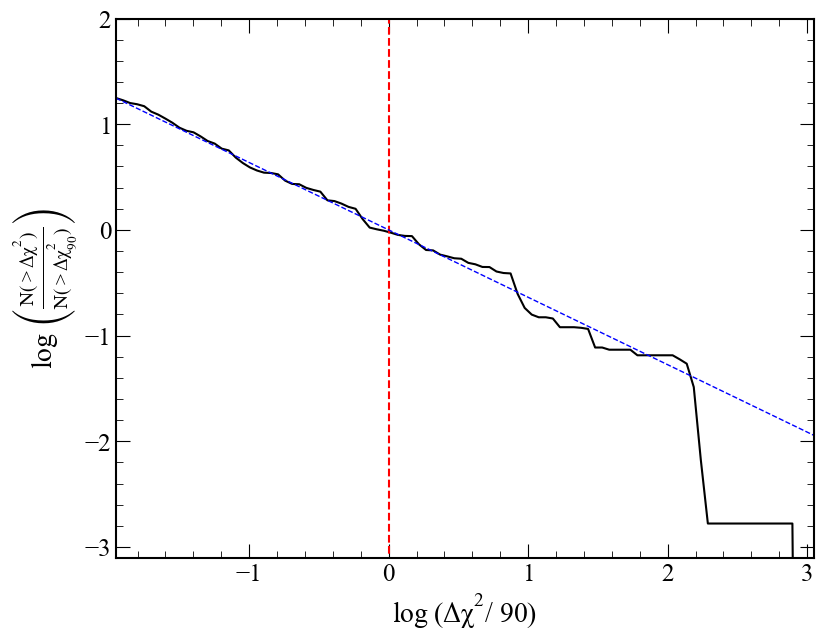}
    \caption{The cumulative distribution of the number $N_{tot}(>\Delta\chi^2)$ of detected moons using a given $\Delta \chi^2_{thr,m}$, normalized to the number of detected moons using $\Delta \chi^2_{thr,m}=90$, is shown as a the black line. The red dashed line is the location of the adopted $\Delta \chi^2_{thr,m} = 90$ threshold. The value of $N_{tot}$ where this red line crosses the black line in our predicted yield of $N_{tot,90}=0.488$. The blue curve shows a power-law fit to the 
    cumulative distribution between $\log \left(\Delta\chi^2/90 \right) = -2$ and $1$. The slope of the fit is $-0.638$.}
    \label{fig:chi2 cumul}
\end{figure}

Summing over all of the detected moons, we find a total yield of $N_{tot}\sim 0.5$ moon detections. This is calculated by summing the weights, $w_i$, of each detected moon, then applying normalization factors described in Section~\ref{sec:romansim}. 

Figure \ref{fig:chi2 cumul} shows the cumulative distribution of the $\Delta \chi^2$ values for the simulated events, i.e., the number of detected events with $\Delta\chi^2$ greater than a given threshold $\Delta \chi^2_{thr,m}$. Our fiducial $\Delta \chi^2_{thr,m} = 90$ threshold is shown; this corresponds a $N_{tot,90}=0.488$ detected planets.  One noteworthy aspect of this plot is that the distribution is approximately linear, such that cumulative distribution of $\Delta \chi^2$ values for moon perturbations approximately follow a power law which we parameterize as
\begin{equation}
   N_{tot,\Delta\chi^2_{thr,m}} = N_{tot,90}\left(\frac{\Delta\chi^2_{thr,m}}{\Delta\chi^2_{90}}\right)^{-\alpha},
    \label{eqn:ndchi2}
\end{equation}
where $N_{tot,\Delta\chi^2_{thr}}$ is the number of detections with $\Delta\chi^2>\Delta\chi^2_{thr}$ and $\alpha$ is the slope.  

This power-law form describes the distribution well until finite source effects become important, at which point the distribution deviates significantly from a simple power law. In the distribution in Figure \ref{fig:chi2 cumul}, this point occurs near $\log \left(\Delta \chi^2_{thr,m} /90 \right) \sim 1$. We fit a linear relationship between $\log \left( \Delta \chi^2_{thr,m} / 90 \right)=-2$ and $\log \left( \Delta \chi^2_{thr,m} / 90 \right)=1$.  We find that $\alpha=0.638$. 

Equation \ref{eqn:ndchi2} can be used to predict how the moon yield would change for different adopted $\Delta \chi^2_{thr}$.  It can also be used to predict how the moon yield would change at fixed $\Delta \chi^2_{thr,m}$ when any parameter that affects $\Delta\chi^2$ is changed, such as the photometric uncertainty or cadence. Specifically,
\begin{equation}\label{eq:new_yield}
    N_{tot,new} \approx N_{tot,old}\left(\frac{\Delta \chi^2_{new}}{\Delta \chi^2_{old}}\right)^{\alpha}.
\end{equation}
As an example, in Section \ref{subsec:higher sampling} we examine how the yield of exomoons changes for shorter survey cadences.

Planetary perturbations from bound planets also exhibit a power-law distribution in $\Delta\chi^2$.  \citet{Penny_2019} examined how the power-law slopes changed for different planet masses. We find a steeper slope ($\alpha=0.638$) than the slope for $0.1 M_{\oplus}$ bound planets ($\alpha=0.534$) using the Roman Cycle 7 survey design found by \citep{Penny_2019}. The typical mass of a simulated moon is $\sim 0.4 M_{\oplus}$. Therefore, we would expect to find a shallower, not steeper, power law distribution. Understanding this discrepancy requires additional work, but it may be due to the interaction between the planet and moon perturbation, such that the presence of the planet suppresses the perturbations created by an isolated moon.

\section{Discussion} \label{sec:discussion}

\subsection{Exomoon Detection Channels} \label{subsec:channels}

We find two possible channels for exomoon detection. The most likely channel is a caustic perturbation, when the source passes over or near the lunar caustic. Nearly 60\% of the detections in our simulations are a result of this channel. Examples are shown in Figures \ref{fig:lc b}a,b,d,e,f,g. The second possible channel is through the deformation of the planetary caustic. The presence of the moon causes slight changes to the planetary caustic that cannot be reproduced by a binary lens model. Although these changes are small, we find that they can be significant, especially if the source crosses the caustic more than twice. An example of this is shown in Figure \ref{fig:lc b}c,h.  This channel is analogous to the detection of circumbinary planets or planets orbiting one member of a wide binary via the distortions they create on the binary lens caustics (see, e.g., \citealt{Gould:2014}).

\subsection{The Effect of Increasing the Sampling Rate}\label{subsec:higher sampling}

Through our work, we find some detections where all of the $\Delta \chi^2$ significance appears to come from a single data point (see Figure \ref{fig:lc b}h). These events would never be claimed as moon detection but are nevertheless significant in that they cannot be reproduced using a planet-only light curve. It is possible, however, that increasing the sampling rate could result in a sufficient number of deviant data points to claim a detection. To test this, we resampled the light curves for the events where an exomoon was detected with sampling rates of ${\cal R}=n{\cal R}_{0}$ with $n=2,3,10$, where ${\cal R}_{0}=(15~{\rm min})^{-1}$ is the sampling rate used in \citet{Penny_2019}.  When doing so, we keep the photometric uncertainties the same.  Thus these are analogous to keeping the exposure time per sample fixed, but decreasing the number of fields monitored in order to decrease the cadence per field. These sampling rates correspond to cadences of 15~min, 7.5~min, 5~min, and 1.5~min. We then fit a binary lens model to these resampled light curves and calculate the $\Delta \chi^2$ for each case. 

\begin{figure}
    \centering
    \includegraphics[width=0.47\textwidth]{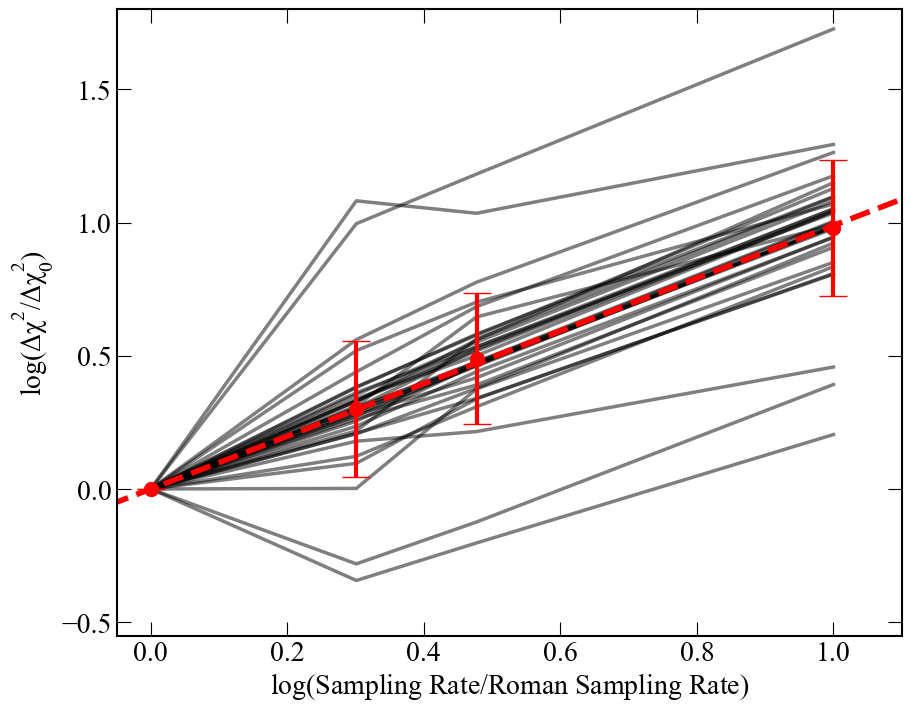}
    \caption{Improvement in $\Delta \chi^2$ of light curves with detected moons as a function of the sampling rate.  The horizontal axis is normalized to the nominal \citet{Penny_2019} sample rate of 15 minutes, whereas the vertical axis is normalized to the $\Delta\chi^2$ value for this sampling rate. The red points are the mean values at each sampling rate with error bars given by the standard deviation of the points in each bin. The red dashed line is the best-fit line to the mean values. We find a slope of $0.99 \pm 0.01$, indicating that the $\Delta \chi^2$ increases linearly with sampling rate, as expected for events for which the majority of the perturbations are well-resolved at the lowest sampling rate.}
    \label{fig:high_sampling}
\end{figure}

Figure \ref{fig:high_sampling} shows the improvement in $\Delta \chi^2$ for each event as a function of the sampling rate increase. The scaling of the improvement in $\Delta \chi^2$ with sampling rate appears similar across most of the events. To quantify this scaling, we bin the $\Delta \chi^2$ values at each sampling rate, determine the standard deviation of the values in the bin, and fit a linear model to $\log{\Delta\chi^2/\Delta\chi^2_0}$ as a function of $\log{{\cal R}/{\cal R}_{0}}$.  We find the best-fit slope of $0.99 \pm 0.01$, consistent with unity.  Thus the $\Delta \chi^2$ increases linearly with the sampling rate. This implies that, for most of the detections, the perturbations are well sampled at the nominal 15-minute cadence. If the perturbations were not well sampled at this rate, we would expect superlinear improvement with sampling rate.  We see that this is the case for two events, for which doubling the sampling rate improves the $\Delta\chi^2$ by a factor of $\sim 10$.  These are events that have only one significantly deviant point at the 15-minute cadence. We also note that increasing the sampling rate {\it decreases} the $\Delta \chi^2$ for $\sim 5\%$ of the detection events.  These are likely artifacts arising from the binary fit minimization routine failing to find the true minimum for the 15-minute cadence. 

Given this scaling of $\Delta \chi^2 \propto {\cal R}$, we can estimate the increase in the yield with increased sampling rate assuming all other survey parameters are fixed using Equation \ref{eq:new_yield},
\begin{equation}\label{eq:calc_new_yield}
    \frac{N_{tot,{\cal R}}}{N_{tot,{\cal R}_0}} \simeq \left[\frac{\Delta \chi^2_{new}({\cal R})}{\Delta \chi^2_{old}({\cal R}_0)}\right]^\alpha \simeq \left(\frac{{\cal R}}{{\cal R}_0}\right)^\alpha = n^{\alpha}.
\end{equation}
If we assume a factor of 10 increase in the cadence (90 seconds), $n=10$, then the yield increases by $\simeq 10^{0.662}\simeq 4.6$.  

Therefore, increasing the cadence by a factor of 10 would lead to a factor of nearly 5 increase in the exomoon yield. We emphasize that this assumes that all other parameters are fixed, including the exposure time and thus measurement uncertainties.  Clearly, this is not possible, as increasing the cadence for a fixed survey area (number of fields) means decreasing the exposure time per field.  However, the factor with which the exposure time decreases is not inversely proportional to $n$ because of the finite slew-and-settle time of Roman.  Nevertheless, with a knowledge of the number and location of the fields and the slew-and-settle time, these approximations can be used to provide a rough estimate of how the number of detected exomoons changes with different sampling rates.

\subsection{Simulating Multiple Moons} \label{subsec:multi moons}

Although we assumed that each planet hosted one moon, systems of multiple moons may be common, particularly in the case where the moons formed in a circumplanetary disk.  For example, in the simulations of \citet{Canup_2006}, the median final number of large moons with $q_m >10^{-5}$ was $N_{lg}=4$. In our Solar System, both Jupiter and Uranus have 4 such moons, although Saturn and Neptune only have one each.  For the same reason that the planetary caustic perturbations in multiple-planet systems are typically well approximated by their linear superposition \citep{Han:2001,Han:2008}, we might expect the effect of multiple moons to be independent of each other and their perturbations to add linearly.  If so, then we can estimate the yields assuming $N_{\lg}$ large moons per system distributed in the same way as the moons we simulated simply by multiplying by $N_{lg}$.  However, just as one moon orbiting a bound planet exhibits more complex behavior than a simple superposition in some regimes (see Section \ref{sec:general}), multiple moons may also exhibit more complex behavior in which the presence of other moons may suppress or enhance the perturbation due to one moon. 

\begin{figure*}
    \centering
    \includegraphics[width=0.9\linewidth]{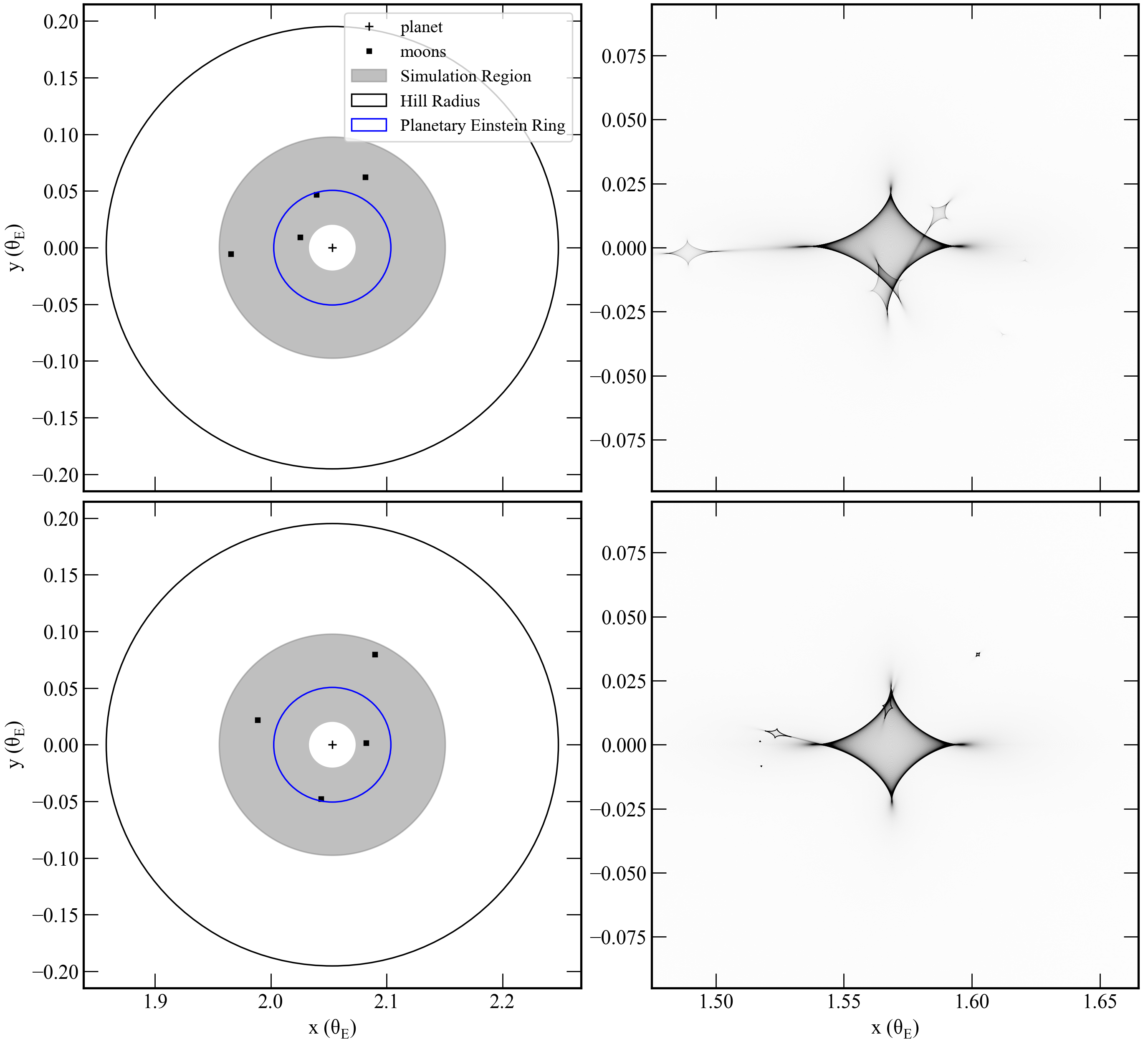}
    \caption{Magnification maps for planetary systems with four moons. In both cases, the host star has a mass of $M=0.37~M_\odot$ and is located at a distance of $D_l=7.9~{\rm kpc}$, and has an Einstein ring radius of $R_{\rm E}=3.0~{\rm AU}$.  The planet has a mass of $M_p=M_{Jup}$ and mass ratio $q_p=2.6\times10^{-3}$, and an instantaneous projected separation of  $6.2~{\rm AU}$ or $s_p=2.08$.  The Hill radius is $a_{\rm Hill}=0.59~{\rm AU}$. The ratio of the Hill radius to the planetary Einstein ring radius is $a_{\rm Hill}/(q_p^{1/2}R_{\rm E})=3.9$.  In each case, the planet hosts four moons with semimajor axes of 0.15, 0.25, 0.35, and 0.45 times the Hill radius.  The moons have the same mass and coplanar orbits with an inclination of $i=45^\circ$.  The phases of the orbits of the moons were chosen at random.   The left panels show the locations of the planet and each moon as black dots, as well as the size of the Hill radius as the black circle and the planetary Einstein Ring as the blue circle. The gray shaded area is the region from 0.1 to 0.5 of the Hill radius in which we simulated moons. The right panels show the resulting magnification maps. In the top row, the mass ratio of each moon is $q_m=10^{-2}$ ($M_m\simeq 3.2~M_\oplus$), and in the bottom row it is $10^{-3}$ ($M_m=0.32~M_{\oplus}\simeq 3~M_{\rm Mars}$.)}
    \label{fig:4moons}
\end{figure*}

To illustrate the effects of multiple moons, in Figure \ref{fig:4moons} we show magnification maps for two systems where the planet hosts four moons. We assume a fixed mass ratio for all the moons of $q_m=10^{-3}$ and $q_m=10^{-4}$ for the top and bottom sets of panels, respectively.  In both cases, we assume that the orbits of the moons are coplanar.  In general, a larger area of the source plane around the planetary caustic is covered with lunar caustics in the case of 4 moons, increasing the probability that a source that passes near the planetary caustic will also pass near a lunar caustic.  The increased complexity of the caustics and magnification maps is also readily apparent in these figures.  While it is difficult to draw any general conclusions from these figures, they do motivate studies of the caustic and light curve phenomenology, as well as the yields, of multiple moon systems.

\subsection{Exclusion of Resonant \& Central Caustic Events} \label{subsec:resonant events}

As discussed in Section \ref{sec:romansim}, we exclude events with resonant caustic configurations from our simulations. However, we expect that exomoons could be detectable in these configurations. Therefore, our predicted yield is a lower limit on the number of exomoons we predict Roman will find. We can estimate the additional yield of moons from resonant caustic events by assuming the detection efficiency is similar across all events. Here, we define the detection efficiency to be, 
\begin{equation}
    {\cal E} = \frac{\sum w_{i,det}}{\sum w_{i,all}},
\end{equation}
where $w_{i,det}$ are the weights of detected events and $w_{i,all}$ are the weights of all simulated 
events, i.e., events that do not include resonant caustics. Multiplying ${\cal E}$ by the yield of all resonant caustic perturbations gives us the number of resonant cases exomoon detections we expect. We determine the yield of all resonant caustic events by summing the weights for all events from GULLS that we excluded from this simulation because they fell in the resonant caustic regime. When we compute ${\cal E}$ considering all non-resonant events (both close and wide), we predict an additional yield of $\sim 0.1$ exomoons. If we only consider wide events when estimating ${\cal E}$, we predict an additional yield of $\sim 0.2$ exomoons. The assumption that the detection efficiencies are the same for resonant events is not well justified, and therefore we urge caution when using these estimates. Simulations of the detectability of moon perturbations in resonant caustic events will be required to predict this yield accurately.

We also exclude events in which the planet was detected due to a central caustic perturbation. Unlike resonant configurations, we do not expect exomoons to be significantly detectable in this configuration. While the size of the planetary caustic is $\propto q_p^{1/2}$, the size of the central caustic is $\propto q_p$ \citep{Chung2005}. The small mass ratios of exomoons would not contribute significantly to the shape of the central caustic, which would be dominated by the planet. Therefore, we do not expect exomoons to be detectable in central caustic events. However, further study is required to more accurately predict this detectability.

\subsection{Orbital Motion of Moons}\label{sec:orbitalmotion}

Here we have ignored orbital motion of the moons about their host planets.  In fact, the moon will move in its orbit over the duration of the perturbation due to the moon.  We can estimate the importance of this effect by evaluating the fraction of its orbit the moon travels during the typical time scale $\Delta t_m$ of the perturbation, $f_{orb} \equiv \Delta t_m/P_m$.  

Typically, the angular Einstein ring of the moon $\theta_{{\rm E},m}$, which sets the scale of the region of the source plane where the perturbations due to the moon are significant, is smaller than the angular radius of the source $\theta_*$.  In this case, the duration of the exomoon perturbation is set by the crossing time of the source, which is $\sim 2\theta_*/\mu_{\rm rel}$.  Therefore the duration of the exomoon perturbation relative the orbital period of the moon is
\begin{equation}
    f_{orb} = \frac{\Delta t_m}{P_m} \simeq \frac{2 \theta_*}{\mu_{\rm rel}P_m} = \frac{2\sqrt{3}\theta_*}{\mu_{\rm rel}P_p}\left(\frac{a_m}{a_{\rm Hill}}\right)^{-3/2},
    \label{eqn:forb}
\end{equation}
where $P_p$ is the period of the planet about the star.  The latter equality follows from the fact that the period of a satellite around a planet with semimajor axis equal to the Hill radius of the planet is equal to the period of the planet around the star divided by $\sqrt{3}$.  

A typical exomoon detected in our simulations is around a planet with a semimajor axis $a_p=4$ AU orbiting a $0.3~M_\odot$ host, which has an orbital period of $P_p\simeq 14.6~{\rm yr}$.  For a turnoff main-sequence source in the Galactic bulge, $\theta_* \sim 0.6~\mu{\rm as}$, and a typical lens-source relative proper motion is $\mu_{\rm rel}\sim 5~{\rm mas~yr^{-1}}$, for a source crossing time of $\theta_*/\mu_{\rm rel} \sim 1~{\rm hr}$.  Thus $f_{orb}$ is typically of order
\begin{equation}
    f_{orb} \sim 10^{-3}\left(\frac{a_p/a_{\rm Hill}}{0.1}\right)^{-3/2}
    \label{eqn:forbeval},
\end{equation}
for a turn-off source and would be an order of magnitude larger for a clump giant source.  

Given the relatively weak nature of the perturbations due to exomoons, we conclude that the effects of orbital motion of the moons are likely negligible.

\section{Conclusions} \label{sec:conclusions}

We conducted simulations of the RGES to predict Roman's ability to detect exomoons of giant planets.  We specifically considered moons orbiting giant planets with masses between $30 M_{\oplus}$ to $10 M_J$ drawn from a a \citet{Cassan_2012} planet mass distribution, with  semimajor axes drawn from a log-uniform distribution from $0.3$ to $30$ AU. We populate these planets with moons with mass ratios $q_m$ in the range $10^{-2}-10^{-4}$ and semimajor axes in the range $[0.1-0.5]a_{\rm Hill}$.  

Our primary conclusions are as follows:
\begin{itemize}
    \item Massive exomoons can be detected using Roman.
    \item Over the course of the Roman Galactic Exoplanet Survey, we expect to find of order one exomoon, given our assumption of one moon per exoplanet with the distribution of mass ratios and semimajor axes we have assumed.  
    \item This yield could be increased with a longer survey, high sampling rate, or if planets typically host multiple massive moons.
    \item Exomoon detections have a variety of phenomenologies, including 'buried' detections perturbations where the source size is significantly larger than the exomoon caustic, moons detected around planets in both close and wide caustic configurations, and cases where the source does not pass over the caustic due to the moon, but rather the moon is detected because it distorts the planetary caustic in a way that cannot be reproduced by a binary lens.
    \item We find that the cumulative distribution of $\Delta\chi^2$ values between the exomoon and planet-only fits to the simulated light curves follows a power-law distribution with a slope that is steeper than that found from analogous simulations of isolated planetary companions with similar mass as the exomoons considered here, implying that the presence of a planet typically suppresses the deviations due to a moon.
    \item We found that a linear increase in the sampling rate leads to a linear increase in the $\Delta\chi^2$ values, implying that the moon perturbations are typically well resolved at our fiducial 15-minute cadence.  
\end{itemize}

More work is needed before we can fully assess Roman's ability to detect and characterize exomoons.  First, we need a more robust and efficient way to quickly and reliably calculate light curves with finite source effects arising from systems of three or more lenses ($N_l \ge 3$) than the code developed used here.  The newly available {\tt VBMicrolensing} code \citep{Bozza:2018} is capable of computing such light curves for an arbitrary $N_l$, although it is not clear how robust it is for the very low mass ratios ($\sim 10^{-8}$) relevant here. Additionally, we must develop algorithms to efficiently fit light curves to extract the moon parameters. We only considered whether the light curve deviated significantly from a planet-only model; we did not attempt to extract the parameters of the moon from the light curves. Fitting planetary microlensing light curves is computationally intensive, and the additional free parameters from the moon will make it even more challenging. Microlensing events also often suffer from degeneracies. We need to determine what degeneracies exist in 3-body and higher-order lenses, how severe these degeneracies are, and how they might be broken. 

The moon systems we simulated are clearly not representative of every exomoon system. To determine Roman's full sensitivity to exomoons, simulations such as the ones we have performed here should be repeated for different types of moons, including analogs to our own Earth-Moon system. The mass of our Moon ($\sim 0.012~M_\oplus$) is generally at the lower end of the masses of the moons we have simulated, and thus analogs to our Moon generally produce less significant signals than the typical moon in our simulation.  However, the Moon/Earth mass ratio is at the upper end of the moon/planet mass ratios we considered, our Moon has a larger value of $a/a_{\rm Hill}$ than the typical moon we simulated, and Earth-mass planets are much more common than giant planets, at least within 1 AU \citep[e.g.,][]{Howard_2012, Fressin_2013}.  Thus the yield of Earth-Moon analogs may be significant depending on how common moon-forming impacts are.

%%\vspace{5mm}
%%\facilities{HST(STIS), Swift(XRT and UVOT), AAVSO, CTIO:1.3m,
%%CTIO:1.5m,CXO}


\software{
{\tt astropy} \citep{exoplanet:astropy13,exoplanet:astropy18}, 
{\tt Jupyter} \citep{jupyer}, 
{\tt Matplotlib} \citep{matplotlib},
{\tt NumPy} \citep{numpy}, 
{\tt SciPy} \citep{scipy}, 
{\tt MulensModel} \citep{Poleski:2019},
{\tt triplelens} \citep{Kuang_2021},
{\tt GULLS} \citep{Penny_2013}
}

\begin{acknowledgments}
We would like to thank the referee for comments and suggestions that significantly improved the paper.
We would like to thank David Kipping for helpful discussions. M.L. would like to thank the Ohio State Department of Astronomy Summer Undergraduate Research Program. M.L. and B.S.G. were supported by the Thomas Jefferson Chair for Space Exploration endowment from the Ohio State University.  B.S.G., M.T.P., and S.A.J. were supported by NASA Grant 80NSSC24M0022.  S.A.J.’s research was supported by an appointment to the NASA Postdoctoral Program at the NASA Jet Propulsion Laboratory, administered by Oak Ridge Associated Universities under contract with NASA.
\end{acknowledgments}

\appendix

\section{Caustic Alignment vs. Anti-Alignment}\label{app:aligned} %get better title later

\begin{figure}
    \centering
    \includegraphics[width=0.9\linewidth]{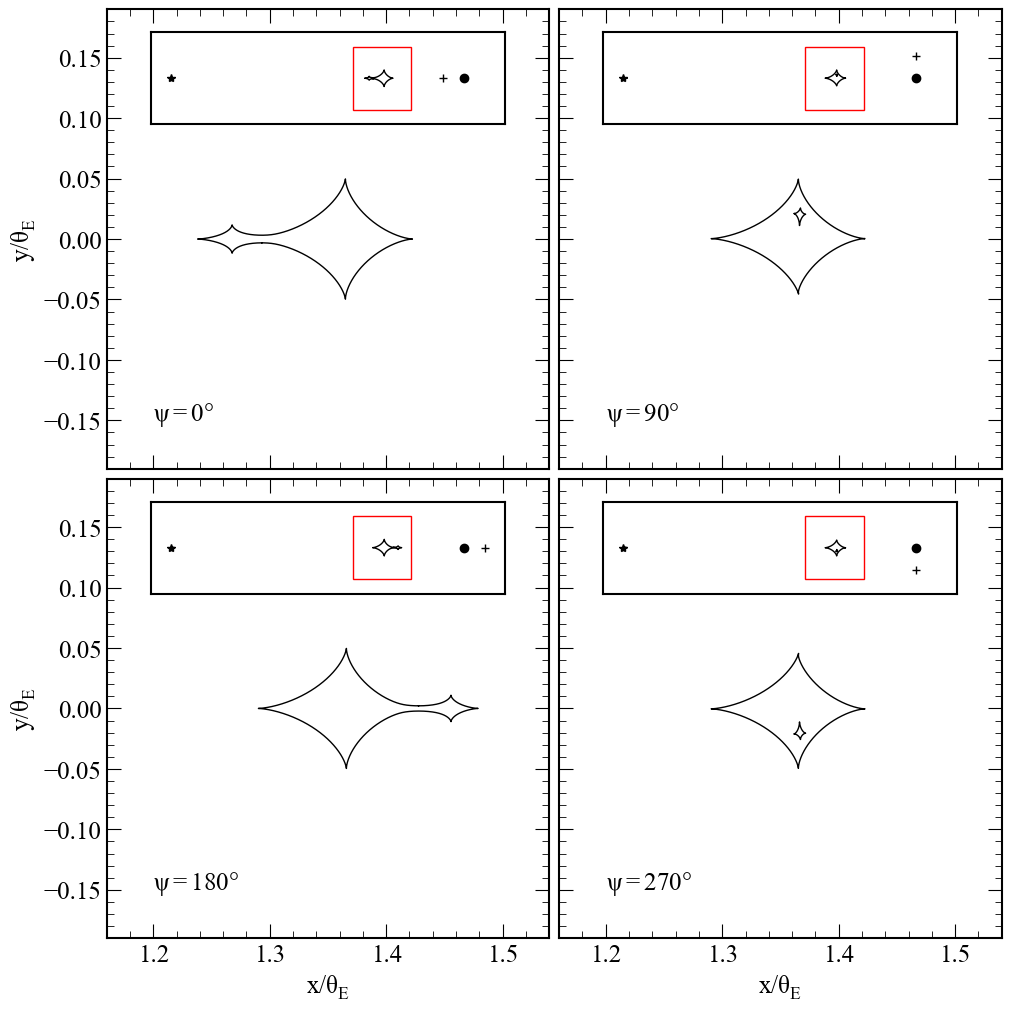}
    \caption{Change in the caustic structure when varying $\Psi$. All other parameters are the same as the light curve in Figure \ref{fig:lc b}a. The inset plot shows the orientation of the star, planet, and moon, and the location of the caustic structures. The star symbol gives the location of the star, the circle denotes the planet's location, and the plus symbol denotes the moon's location. Changing $\Psi$ changes the orientation of the moon around the planet. The red box denotes the area of the map shown in the main part of the plot.  This figure shows how the position and size of the lunar caustic changes with different orientations of the moon. It is clear that the lunar caustic is smaller when the moon-planet axis is perpendicular to the star-planet axis ($\Psi = 90 \degr, 270\ \degr$) than when it is aligned ($\Psi = 0\degr, 180\degr$), and thus these configurations when the moon is aligned are more likely to be detected.}
    \label{fig:enter-label}
\end{figure}

In Section \ref{sec:results} and Figure \ref{fig:histograms}, we noted that moons were preferentially detected at angles of $\Psi \simeq 0^\circ$ and $\Psi=180^\circ$, i.e. when the moon is aligned or anti-aligned with the planet-star axis.  In order to explore the reason for this preference, we consider the caustics produced by a moon for an example case when the shear due to the star on the moon is comparable to the shear to the planet on the moon.  Figure \ref{fig:enter-label} shows the caustic structures for the same physical parameters of the system shown in Figure \ref{fig:lc b}a, but varying $\Psi$ at $0 \degr$, $90 \degr$, $180 \degr$, and $270 \degr$.  For this example, $\gamma_s=1/s_p^2=0.28$ and $\gamma_p=1/s_m^2=0.58$, and thus $\gamma_p/\gamma_s\sim 2$.  We see that the lunar caustic is smaller when the moon-planet axis is anti-aligned with the star-planet axis ($\Psi = 90\degr, 270\degr$) than when it is aligned with the star-planet axis ($\Psi = 0\degr, 180\degr$). We suspect that this is likely due to interference from the star and the planet. When the moon is anti-aligned, the shear due to the star destructively interferes with the shear due to the planet, whereas when the moon is aligned, the shears constructively interfere. A similar effect is seen in the lensing of a quasar due to an elliptical galaxy with an external perturber when the major axis of the elliptical galaxy is orthogonal to the direction of the perturber \citep{Keeton:2000}. 

The size of the caustic produced by the moon affects its detectability.  In general, larger caustics are easier to detect, since the probability that a source passes near a caustic is directly proportional to its cross-section. Therefore, a moon that produces a larger caustic will have a higher detection probability. Since the lunar caustic is larger when the moon is aligned with the planet-star axis, we expect the moon to be more detectable when it is aligned than when it is anti-aligned, at least when $\gamma_s\sim \gamma_p$.  Given that, for moons with orbits that are coplanar with the planet orbit, angles of $\Psi \sim 0$ and $\Psi\sim 180\degr$ are also {\it a priori} more likely, this effect is likely to boost the detectability of moons. 

\section{Table of Length Variables}

\begin{table}
\centering
\caption{Table of Length Scales used in this paper. Each row lists the variable assigned to that length scale, a short definition, the dimension of the variable (whether it is a physical length or  dimensionless), the units used for that variable if it has dimension, and the normalization reference value if it is dimensionless.}
\label{tab:variables}
\begin{tabular}{c|>{\centering\arraybackslash}p{9.5cm}|c|c|c}
\hline
\textbf{Variable} & \textbf{Definition} & \textbf{Dimension} & \textbf{Unit} & \textbf{Normalization} \\
\hline
$D_l$ & distance to the lens & length & pc, kpc & - \\
$D_s$ & distance to the source & length & pc, kpc & - \\
$a_m$ & semimajor axis of the moon to the planet & length & AU & - \\
$a_p$ & semimajor axis of the planet to the star & length & AU & - \\
$a_{\rm Hill}$ & Hill radius of the planet & length & AU & - \\
$R_{E,p}$ & Einstein ring radius of the planet & length & AU & - \\
$\zeta$ & positions of the source & dimensionless & - & $\theta_E$\\
$z$ & positions of images & dimensionless & - & $\theta_E$ \\
$x_{cen}$ & distance between the star and the center of the planetary caustic & dimensionless & - & $\theta_E$ \\
$u$ & angular separation between the source and lens & dimensionless & - & $\theta_E$ \\
$u_0$ & impact parameter of the source & dimensionless & - & $\theta_E$ \\
$\rho_*$ & angular radius of the source star & dimensionless & - & $\theta_E$ \\
$s_c$ & boundary between close and resonant caustic regimes & dimensionless & - & $\theta_E$ \\
$s_w$ & boundary between resonant and wide caustic regimes & dimensionless & - & $\theta_E$ \\
$s_{caus}$ & cross-section of the planetary caustic & dimensionless & - & $\theta_E$ \\
$\Delta \eta$ & horizontal width of the planetary caustic & dimensionless & - & $\theta_E$ \\
$\Delta \xi$ & vertical width of the planetary caustic & dimensionless & - & $\theta_E$ \\
$s_p$ & instantaneous projected separation between the lens star and planet & dimensionless & - & $\theta_E$ \\
$s_m$ & instantaneous projected separation between the planet and moon & dimensionless & - & $\theta_{E,p}$ \\
$s_{\rm Hill}$ & ratio of the Hill radius of the planet to the planetary Einstein ring radius & dimensionless & - & $R_{E,p}$ \\
\hline
\end{tabular}
\end{table}

Throughout this paper, we define many different length scales, many of which have different units. To avoid confusion, we list all variables referring to a length scale in Table \ref{tab:variables} along with a short definition of that length scale. It also lists whether the scales has physical units or is dimensionless and its corresponding units or normalization, respectively.

\bibliography{references}{}
\bibliographystyle{aasjournal}

\end{document}